\documentclass[
  reprint,
  superscriptaddress,
  amsmath,
  amssymb,
  aps,
  prx,
  floatfix,
  raggedbottom,
  longbibliography
]{revtex4-2}

\usepackage{bm}
\usepackage{booktabs}
\usepackage{graphicx}
\usepackage{mathtools}
\usepackage{xcolor}
\usepackage[colorlinks=true,citecolor=blue,linkcolor=black,urlcolor=black]{hyperref}

\hypersetup{pdftitle={SMP: A General Hyperedge-Based Framework for Circuit-Level Quantum Error Correction},pdfauthor={Siying Wang et al.}}
\graphicspath{{figures/}}
\newcommand{\ColorRows}{%
0.001 & 9 & X & 0.306 & 0.165 & 46.1 & $2.6\times10^{-32}$ \\
0.001 & 9 & Z & 0.258 & 0.142 & 45.0 & $7.2\times10^{-28}$ \\
0.001 & 11 & X & 0.102 & 0.048 & 52.9 & $2.7\times10^{-12}$ \\
0.001 & 11 & Z & 0.107 & 0.049 & 54.2 & $2.1\times10^{-14}$ \\
0.001 & 13 & X & 0.038 & 0.010 & 73.7 & $7.7\times10^{-7}$ \\
0.001 & 13 & Z & 0.036 & 0.008 & 77.8 & $7.5\times10^{-9}$ \\
0.003 & 9 & X & 7.192 & 5.092 & 29.2 & $<10^{-300}$ \\
0.003 & 9 & Z & 7.025 & 4.989 & 29.0 & $<10^{-300}$ \\
0.003 & 11 & X & 5.668 & 3.594 & 36.6 & $<10^{-300}$ \\
0.003 & 11 & Z & 5.617 & 3.642 & 35.2 & $<10^{-300}$ \\
0.003 & 13 & X & 4.356 & 2.441 & 44.0 & $<10^{-300}$ \\
0.003 & 13 & Z & 4.339 & 2.455 & 43.4 & $<10^{-300}$ \\%
}
\newcommand{\GateRows}{%
LOM + MWPM & 0.535 & [0.443, 0.646] & --- \\
SMP + LOM + MWPM & 0.345 & [0.273, 0.436] & 35.5 \\
BP10-assisted LOM & 0.320 & [0.251, 0.408] & 40.2 \\
BP100-assisted LOM & 0.260 & [0.198, 0.341] & 51.4 \\%
}
\newcommand{\ColorReductionRows}{%
9 & 46.1 & 45.0 & 29.2 & 29.0 \\
11 & 52.9 & 54.2 & 36.6 & 35.2 \\
13 & 73.7 & 77.8 & 44.0 & 43.4 \\%
}
\newcommand{\GateMainRows}{%
LOM & 0.535 & [0.443, 0.646] & --- \\
SMP + LOM & 0.345 & [0.273, 0.436] & 35.5 \\
BP10-assisted LOM & 0.320 & [0.251, 0.408] & 40.2 \\%
}

\begin{document}

\title{SMP: A General Hyperedge-Based Framework for Circuit-Level Quantum Error Correction}

\author{Siying Wang}
\thanks{These authors contributed equally to this work.}

\author{Yue Yan}
\thanks{These authors contributed equally to this work.}

\author{Zhixin Xia}

\author{Canwei Shi}
\author{Hannuo Yuan}
\affiliation{State Key Laboratory of Low Dimensional Quantum Physics, Department of Physics, Tsinghua University, Beijing 100084, China}

\author{Xiang-Bin Wang}
\email{xbwang@mail.tsinghua.edu.cn}

\affiliation{State Key Laboratory of Low Dimensional Quantum Physics, Department of Physics, Tsinghua University, Beijing 100084, China}
\affiliation{Jinan Institute of Quantum Technology and Jinan branch, Hefei National Laboratory, Jinan, Shandong 250101, China}
\affiliation{Frontier Science Center for Quantum Information, Beijing 100193, China}

\date{\today}

\begin{abstract}
Hyperedge faults remain a major challenge in circuit-level quantum error correction. Accurately exploiting hyperedge correlations often requires substantial computational resources, which complicates real-time decoding. Matching-based decoders are fast and scalable, but their pairwise graph representation limits the use of higher-order fault information. Here, we introduce Syndrome Motif Projection (SMP), a general framework for incorporating hyperedge information into matching-based decoding. The key idea is that a hyperedge fault produces a characteristic local syndrome motif. Observing such a motif provides evidence for the corresponding fault. SMP extracts this information directly from the measured syndrome with linear-complexity preprocessing while preserving the matching backend. It can therefore serve as a lightweight front end to existing scalable and real-time matching decoders. We demonstrate SMP across surface code memories, color code memories, and logical gate circuits. On Google's Willow experimental surface code data, SMP improves both standard and correlated matching and outperforms the tested belief matching baseline. For color code memories, SMP combined with Chromobius reduces logical failure probabilities by up to 77.8\%. For logical gate circuits, SMP reduces the failure probability of a six-CNOT circuit by 35.5\% relative to logical observable matching while achieving performance comparable to iterative belief matching. These results demonstrate that SMP provides a lightweight and transferable approach to hyperedge decoding, with a clear path toward integration into scalable real-time decoding architectures for fault-tolerant quantum computation.
\end{abstract}
\maketitle
\section{Introduction}
\label{sec:introduction}
Fault-tolerant quantum computation relies on fast and accurate classical decoding of measured syndromes \cite{fowler2012practical,circuit_2023improved,willow2025,experiment_realization_2022,ni2023breakeven,he2025leakagesuppression}. However, hyperedge faults are a recurring challenge in circuit-level decoding \cite{delfosse2023splitting}. A single circuit fault can trigger three or more correlated detection events, forming a hyperedge in the detector error model \cite{gidney2021stim,circuit_2023improved}. In realistic quantum computing experiments, hyperedge faults can arise from circuit-level fault propagation, correlated noise such as crosstalk \cite{zhou2025crosstalk,liu2025parallelcrosstalk,correlated_crosstalk,detuning_zhao2022quantum}, and fault propagation across logical gate operations \cite{cain2024correlated}, and are also intrinsic to the syndrome structure of several quantum error-correcting codes \cite{sahay2022mobius,serraperalta2026lom,wang2026lowoverhead,liang2025generalizedtoric}. 

Several approaches have been developed to handle hyperedge faults. Neural decoders can learn joint syndrome patterns from data \cite{bausch2024learning,alphaqubit2,yang2026fpgadecoder}. Hypergraph and search-based decoders can also model multi-detector fault mechanisms explicitly \cite{beni2025tesseract}. These approaches can achieve high decoding accuracy, but often require substantial computation or model complexity \cite{bausch2024learning,beni2025tesseract}. Matching-based decoders are a leading practical choice for fault-tolerant quantum error correction. Fast implementations, parallelization, and windowed architectures make them well suited to large-scale and real-time decoding \cite{sparse_blossom,fusion_blossom,tan2023parallel,skoric2023window}.

Existing matching-based decoders still face a trade-off between decoding accuracy and computational cost. Correlated minimum-weight perfect matching (MWPM) uses correlations associated with hyperedge faults to reweight edges in the matching graph \cite{fowler2013correlated,paler2023pipelined}, but remains less accurate than more powerful decoders \cite{bausch2024learning}. Belief matching extracts richer hyperedge information through iterative belief propagation on the circuit-level Tanner graph \cite{circuit_2023improved}, but it requires repeated message passing and depends strongly on the prior fault model \cite{sivak2024priors}. How to exploit hyperedge information more effectively without substantial additional complexity or resource cost therefore remains an important challenge for matching-based decoding.

Here, we introduce Syndrome Motif Projection (SMP), a general framework for handling hyperedge faults in matching-based decoding. The main idea is that a hyperedge fault produces a characteristic local syndrome motif. When nearby detection events match this pattern, they provide evidence that increases the posterior probability of the corresponding hyperedge fault. SMP approximates this local evidence and maps it onto the associated matching edge weights. Unlike belief matching, which performs iterative inference on the Tanner graph \cite{circuit_2023improved}, SMP uses direct local updates to the matching graph. This provides a lightweight way to exploit hyperedge information and can be adapted to different codes and circuit settings.

We demonstrate SMP in three settings: surface code memories, color code memories, and logical gate circuits. For surface codes, SMP projects local hyperedge evidence directly onto the native matching graph. On Google's Willow data, it improves both standard and correlated MWPM under both tested decoder priors \cite{willow2025}. At code distance $d=7$ over 13 extraction rounds, SMP-correlated MWPM also achieves lower observed failure probabilities than the tested belief matching baseline with up to 10 BP iterations. For color codes, SMP transfers motif information to the M\"obius matching graph used by Chromobius \cite{sahay2022mobius,gidney2023chromobius}. It reduces the logical failure probability by up to 77.8\% over the tested range. For logical gate circuits, SMP transfers information from complete motifs across separate logical observable matching graphs \cite{serraperalta2026lom}. In a circuit with six successive transversal CNOTs and one error correction round between gates, SMP reduces the circuit failure probability by 35.5\% relative to logical observable matching. These results suggest that SMP is applicable to a broad range of quantum error correction settings. 

\section{Surface code decoding under circuit-level noise}
\label{sec:background}
\begin{figure*}[tp]
  \centering
  \includegraphics[width=\textwidth]{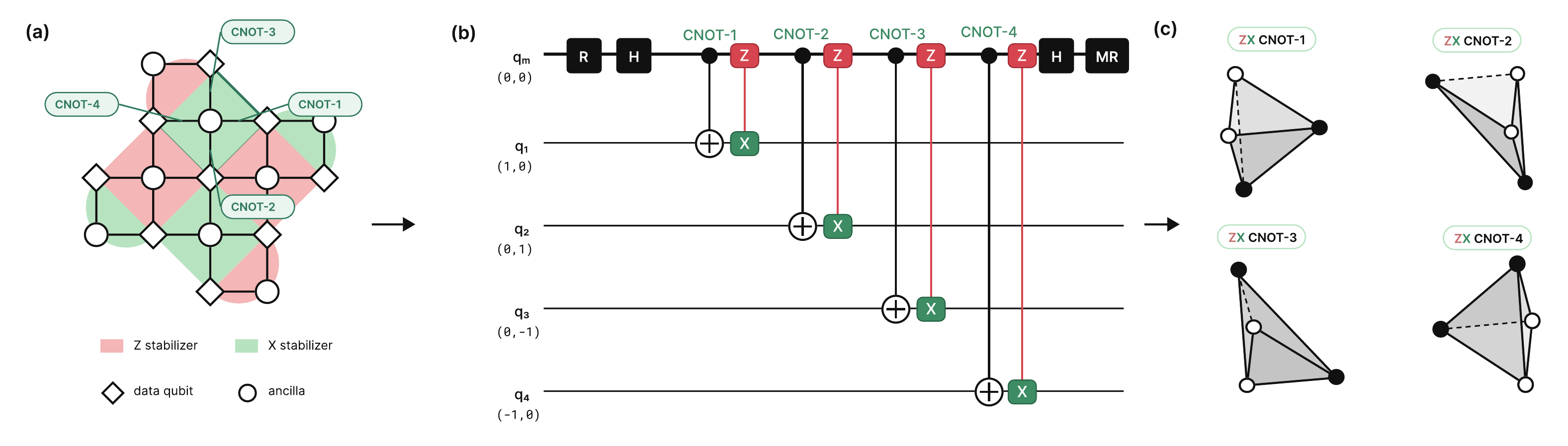}
  \caption{Four-detector hyperedges induced by circuit-level faults. \textbf{(a)} Rotated surface code patch. Green and red plaquettes denote $X$- and $Z$-type stabilizers, respectively; diamonds denote data qubits and circles denote measurement qubits. The four CNOT gates used to measure the highlighted $X$-type stabilizer are labeled CNOT-1 through CNOT-4. \textbf{(b)} Circuit for one syndrome extraction round (round 0). The measurement qubit $q_m$ at spatial coordinate $(0,0)$ interacts sequentially with data qubits $q_1$, $q_2$, $q_3$, and $q_4$ at $(1,0)$, $(0,1)$, $(0,-1)$, and $(-1,0)$, respectively. The four $ZX$ markers are alternative single-fault locations: after CNOT-$i$, a $Z$ error acts on $q_m$ and an $X$ error acts on $q_i$, while the other CNOT gates are fault-free. \textbf{(c)} The resulting four-detector supports, with detector coordinates $(x,y,t)$ defined relative to $q_m$. Temporal propagation produces the common white vertex pair $\mathcal W=\{(0,0,0),(0,0,1)\}$. For CNOT-1 through CNOT-4, the gate-dependent black vertex pairs are, respectively, $\{(+1,+1,0),(+1,-1,0)\}$, $\{(-1,+1,1),(+1,+1,0)\}$, $\{(-1,-1,1),(+1,-1,0)\}$, and $\{(-1,+1,1),(-1,-1,1)\}$; each four-detector support is the union of $\mathcal W$ and the corresponding black vertex pair.}
  \label{fig:circuit-noise-motifs}
\end{figure*}

The surface code encodes logical information in the common $+1$ eigenspace of local $X$- and $Z$-type stabilizers, which generate the stabilizer group $\mathcal G$ \cite{Surface_code2,Surface_code1,surface_code_intro,surface_code_intro2015,gottesman1997stabilizer}. Figure~\ref{fig:circuit-noise-motifs}(a) shows a rotated surface code patch. Measurement qubits interact with neighboring data qubits to measure these stabilizers without measuring the encoded logical state. A Pauli error changes the eigenvalue of each stabilizer with which it anticommutes. The pattern of changed eigenvalues is the error syndrome. A decoder uses this information to infer a recovery operator $R$ for the data error $E$. If $R$ and $E$ have the same syndrome, the residual operator $RE$ commutes with all stabilizers and returns the state to the code space. This alone does not guarantee recovery of the logical state. Up to an overall phase, decoding succeeds when $RE\in\mathcal G$ and fails when $RE\in\mathcal N(\mathcal G)\setminus\mathcal G$. Here $\mathcal N(\mathcal G)$ denotes the Pauli operators that commute with all stabilizers; those outside $\mathcal G$ act nontrivially on the logical state. The same correction principle applies to the XZZX code, with its different stabilizer structure \cite{rotated_horsman2012surface,rotatedcode_2025compare,xzzx_2021}.

Circuit faults can also corrupt stabilizer measurements, so syndrome extraction is repeated over multiple rounds \cite{quantum_codes4,quantum_codes5,experiment_realization_2022}. At the circuit level, depolarizing noise is commonly used to model stochastic faults \cite{surface_code_intro,gidney2021stim,circuit_2023improved}. Let $U_{\mathrm{CNOT}}$ be the ideal CNOT unitary. For an input state $\rho$, let $
\rho_{\mathrm C}=U_{\mathrm{CNOT}}\rho U_{\mathrm{CNOT}}^{\dagger}
$
denote the state after the ideal gate operation. The corresponding noisy channel is
\begin{equation}
\mathcal N_{\mathrm{CNOT}}(\rho)
=(1-p)\rho_{\mathrm C}
+\frac{p}{15}
\sum_{P\in\mathcal P_2\setminus\{I\otimes I\}}
P\rho_{\mathrm C}P ,
\label{eq:circuit-depolarizing-channel}
\end{equation}
where \(\mathcal P_2=\{I,X,Y,Z\}^{\otimes 2}\). Here $p$ is the total probability of a nonidentity two-qubit Pauli error.
These faults then propagate through the remaining syndrome extraction circuit and trigger several detectors at different positions and times. The resulting detector pattern depends not only on the Pauli error, but also on the circuit location where the fault occurs \cite{circuit_2023improved,willow2025,fowler2012autotune}.

Figure~\ref{fig:circuit-noise-motifs}(b) shows one round of \(X\) stabilizer measurement around the central measurement qubit \(q_m\). The four CNOT gates are labeled CNOT 1 to CNOT 4. We consider a \(ZX\) fault immediately after each CNOT and propagate it through the rest of the circuit. Figure~\ref{fig:circuit-noise-motifs}(c) shows the corresponding detector patterns. Although the same \(ZX\) error is considered in each case, faults at different CNOT locations produce different four detector motifs. These motifs therefore contain information about the underlying hyperedge faults, which we exploit for decoding below.

\section{Syndrome Motif Projection}
\label{sec:smp}
\subsection{Local posterior evidence and normalized graph reweighting}
\label{sec:smp_posterior}
As discussed in Sec.~\ref{sec:background}, hyperedge faults leave characteristic syndrome patterns that provide local evidence for their occurrence. The key question is how to incorporate this evidence into a pairwise matching graph. In this subsection, we first develop an explicit local-posterior formulation that quantifies the evidence for a candidate hyperedge and converts it into a bounded modification of the matching weights. This formulation serves as a reference model for SMP. The next subsection introduces SMP as a simpler approximation, while Appendix~\ref{app:local-posterior} implements the explicit local calculation and compares the two approaches.

To evaluate a candidate fault and represent it in the decoder, we need its prior probability, detector support, and logical effect. A detector error model (DEM) supplies these quantities for independent stochastic error mechanisms \cite{gidney2021stim}. We index the mechanisms by $\mathcal R$, denote the detector set by $\mathcal V_{\mathrm{det}}$, and write $n_L$ for the number of tracked logical observables:
\begin{equation}
 \mathrm{DEM}=\bigl\{(p_r,S_r,\bm\ell_r)\bigr\}_{r\in\mathcal R},
 \qquad
 \begin{aligned}
  &p_r\in[0,1],\quad S_r\subseteq\mathcal V_{\mathrm{det}},\\
  &\bm\ell_r\in\mathbb F_2^{n_L}.
 \end{aligned}
 \label{eq:smp_dem_definition}
\end{equation}
Here $p_r$ is the probability that mechanism $r$ occurs. The support $S_r$ contains the detectors flipped by $r$. Each entry of $\bm\ell_r$ records whether $r$ flips the corresponding logical observable (one for a flip, zero otherwise). We focus on faults that flip three or more detectors, which cannot be represented by a single matching edge. These hyperedge faults form the set
\begin{equation}
 \mathcal H=\{r\in\mathcal R:|S_r|\geq3\}.
\end{equation}
For $h\in\mathcal H$, its detector support $S_h$ is called a syndrome motif. Mechanisms with the same support can have different logical-observable flip vectors. When independent mechanisms with identical $(S_r,\bm\ell_r)$ are grouped, we also use $h$ for the resulting effective fault. Its support $S_h$ and logical-observable flip vector $\bm\ell_h$ are unchanged; its prior $p_h$ is given by Eq.~\eqref{eq:smp_combined_prior}.

The matching decoder selects edges to explain the observed detection events. For the surface code, its graph has vertices representing detectors and boundaries. We denote its edge set by $\mathcal E$. Each edge $e\in\mathcal E$ connects either two detectors or one detector to a boundary vertex. Errors that trigger at most two detectors are called \emph{graphlike errors}. Those with one or two detection events are represented by matching edges. Each edge has a prior probability $0<p_e<1$ and a weight $b_e=\ln[(1-p_e)/p_e]$. Its logical-observable flip vector $\bm\ell_e\in\mathbb F_2^{n_L}$ has one entry per tracked logical observable: 1 if the error represented by $e$ flips it, and 0 otherwise. 

To represent a hyperedge fault $h$ in this graph, we need a set of edges with the same detector support $S_h$ and logical-observable flip vector $\bm\ell_h$. Consider an edge set $E\subseteq\mathcal E$. Each edge flips its detector endpoints. Repeated flips at a detector cancel in pairs. The resulting detector support, denoted by $\partial E$, therefore contains exactly the detectors connected to an odd number of edges in $E$. A boundary endpoint has no associated detector outcome. Thus, an edge joining a detector to a boundary contributes only its detector endpoint to $\partial E$. For each logical observable, an odd number of flips gives a net flip, whereas an even number gives no net flip. The combined logical-observable flip vector is therefore
\begin{equation}
 \bm\ell(E)=\bigoplus_{e\in E}\bm\ell_e,
 \label{eq:smp_edge_parities}
\end{equation}
where $\bigoplus$ denotes componentwise addition modulo two. We call an edge set $C_h\subseteq\mathcal E$ a cluster representing $h$ if
\begin{equation}
 \partial C_h=S_h,
 \qquad
 \bm\ell(C_h)=\bm\ell_h.
 \label{eq:smp_cluster_definition}
\end{equation}

At the circuit level, a detector is a predefined parity check on measurement outcomes, represented by a vertex in the decoding graph. A detection event occurs when this parity differs from its fault-free value. We denote the binary detection-event vector by $\bm y$, with $y_d=1$ for a detection event at detector $d$ and $y_d=0$ otherwise. Hereafter, syndrome refers to these detection-event data, rather than the stabilizer eigenvalue changes described in Sec.~\ref{sec:background}. To assess the local evidence for $h$, we compare its candidate motif $S_h$ with the observed event set $Y=\{d\in\mathcal V_{\mathrm{det}}:y_d=1\}$. The observed part of the motif is $O_h=S_h\cap Y$. We denote the observed event count by $n_h=|O_h|$ and the full motif size by $k_h=|S_h|$. A complete motif has $k_h-n_h=0$; a motif with one missing event has $k_h-n_h=1$.

A complete motif or a motif with one missing event ($k_h-n_h\leq1$) can provide strong local evidence for the corresponding hyperedge fault $h$. For example, the surface-code circuit in Fig.~\ref{fig:circuit-noise-motifs}(c) produces a complete motif with $n_h=k_h=4$, denoted $4/4$. Suppose the elementary fault probabilities scale linearly with the noise strength $p$. This motif can arise from $h$ alone with probability $O(p)$. A competing explanation with two independent edge errors has probability $O(p^2)$. If one event is missing, $k_h-n_h=1$ and $(n_h,k_h)=(3,4)$, giving a $3/4$ motif in this example. The hyperedge must then occur together with another fault that cancels that event. This explanation has probability $O(p^2)$, the same order as the two-independent-edge-error explanation.

\begin{figure}[!htbp]
\centering
\includegraphics[width=0.82\columnwidth]{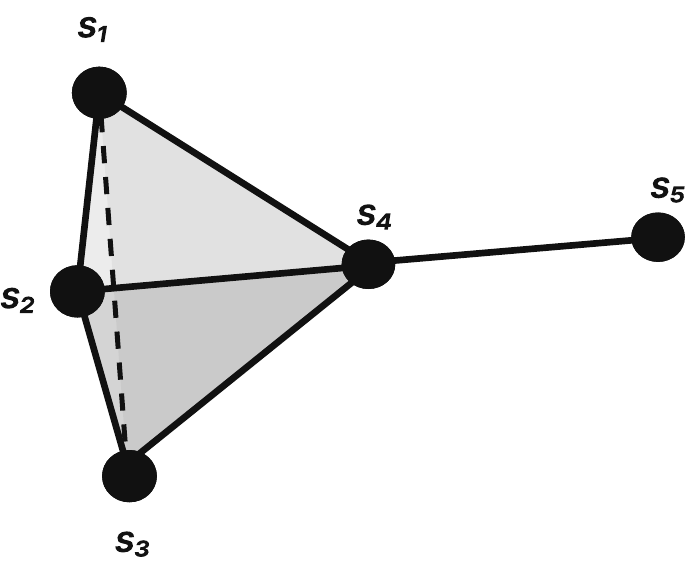}
\caption{Local motif evidence does not determine a complete correction. The labels \(s_1,\ldots,s_5\) denote detector locations with observed detection events. The shaded region highlights a candidate syndrome motif on \(s_1,\ldots,s_4\). Committing those four events to one local correction leaves \(s_5\) to be explained by other events or a boundary, neither of which is shown. The connection between \(s_4\) and \(s_5\) illustrates a competing use of a detector inside the motif. The drawing is schematic and assigns no numerical edge costs.}
\label{fig:local-global}
\end{figure}

However, this local order comparison does not determine whether the hyperedge belongs to the globally optimal correction, which must also explain detection events outside the motif. In Fig.~\ref{fig:local-global}, the four events $s_1,\ldots,s_4$ form a candidate motif. Even if this pattern strongly favors the hyperedge, $s_4$ may instead be paired with $s_5$ through a short edge. Assigning all four events to the hyperedge would leave $s_5$ to be matched elsewhere, which could increase the total cost. We therefore use local evidence to adjust edge weights and let the matching decoder choose the complete correction.

Thus, the detector pattern on $S_h$ admits multiple explanations involving hyperedge faults or graphlike errors. We construct a local graph region containing the candidate motif, overlapping candidate motifs, and selected matching edges incident to their detectors. Let $\Omega\subseteq\mathcal V_{\mathrm{det}}$ denote the detector set of this region. We write $\bm y_\Omega$ for the syndrome restricted to $\Omega$. The conditional probability of $h$ is
\begin{equation}
 \Pr(h\mid\bm y_\Omega)
 =\frac{p_h\Pr(\bm y_\Omega\mid h)}
 {p_h\Pr(\bm y_\Omega\mid h)
 +(1-p_h)\Pr(\bm y_\Omega\mid\neg h)}.
 \label{eq:smp_local_bayes}
\end{equation}
Here, $h$ and $\neg h$ mean that the candidate fault occurs or is absent. Without $h$, independent edge errors must explain the observed events $Y\cap\Omega$. With $h$, they must explain the residual events $(Y\triangle S_h)\cap\Omega$. Here $Y\triangle S_h$ means removing observed events in $S_h$ and adding the missing ones: each detector in $S_h$ is flipped once.

We compare these alternatives by adding the probability weights of their allowed edge configurations. Under the independent-edge approximation, selecting edge $e$ replaces a factor $1-p_e$ by $p_e$. Relative to no edge errors, an edge set $E$ therefore has weight
\begin{equation}
 e^{-W_0(E)}=\prod_{e\in E}\frac{p_e}{1-p_e}.
 \label{eq:smp_local_edge_mass}
\end{equation}
Here $W_0(E)=\sum_{e\in E}b_e$ is the original matching cost. The alternative containing $h$ also includes its prior odds $P_H=p_h/(1-p_h)$. Summing the weights for each alternative gives
\begin{equation}
 \begin{aligned}
  e^{-c_I}&=\sum_E^{\mathrm{local},I}e^{-W_0(E)},\\
  e^{-c_h}&=P_H\sum_E^{\mathrm{local},h}e^{-W_0(E)}.
 \end{aligned}
 \label{eq:smp_local_configuration_evidence}
\end{equation}
The costs $c_I$ and $c_h$ are the negative logarithms of the total weights without and with $h$. The sums use the allowed local configurations specified in Appendix~\ref{app:local-posterior}. For an isolated candidate, dividing its weight by the sum of both weights gives
\begin{equation}
 \alpha_h=\frac{e^{-c_h}}{e^{-c_I}+e^{-c_h}}
 =\frac{\eta_h}{1+\eta_h},
 \qquad \eta_h=e^{-(c_h-c_I)}.
 \label{eq:smp_local_posterior_share}
\end{equation}
We call $\eta_h$ the evidence ratio. It includes the prior odds $P_H$ and is therefore the posterior odds within this two-explanation model, rather than a likelihood ratio alone. The normalized probability $\alpha_h$ approximates Eq.~\eqref{eq:smp_local_bayes}.

Matching needs an additive cost for each edge, whereas $\alpha_h$ describes the complete candidate. We therefore use the physical prior to set the target cluster cost and the local probability to control the change toward that target. The original and target costs are
\begin{equation}
 W_{E,h}=\sum_{e\in C_h}b_e,
 \qquad W_{C,h}=-\ln P_H=\ln\frac{1-p_h}{p_h}.
 \label{eq:smp_posterior_cost_reference}
\end{equation}
We consider $0<p_h<1/2$ and retain proposals with $W_{C,h}<W_{E,h}$. We interpolate the cluster cost as $(1-\alpha_h)W_{E,h}+\alpha_hW_{C,h}$. To implement this change through individual edges, we distribute it in proportion to their original costs:
\begin{equation}
 \begin{aligned}
  w_e
  &=b_e+\alpha_h\frac{b_e}{W_{E,h}}
       (W_{C,h}-W_{E,h})\\
  &=(1-\alpha_h)b_e+\alpha_h\widetilde b_{h,e},
  \qquad e\in C_h,
 \end{aligned}
 \label{eq:smp_posterior_normalized_weight}
\end{equation}
Here $\widetilde b_{h,e}=b_eW_{C,h}/W_{E,h}$ is the proposed edge weight. Summing $w_e$ over $C_h$ gives the interpolated cluster cost.

\subsection{Motif approximation for isolated and overlapping clusters}
\label{sec:smp_approximation}

For each observed syndrome, Eq.~\eqref{eq:smp_local_configuration_evidence} requires finding edge configurations for both the observed events and the residual events after applying the candidate fault. Our aim is to avoid this repeated search by identifying the matching edges associated with each candidate motif once, then directly adjusting their weights using the observed part $O_h$ of the motif.

For each candidate motif $S_h$, we therefore determine a cluster $C_h$ once and use the same edges for subsequent syndrome-dependent weight updates. Let $\mathcal E(S_h)$ contain edges between detectors in $S_h$ and edges from those detectors to a boundary. We choose the lowest-cost local representation that preserves the candidate's detector support $S_h$ and logical effect $\bm\ell_h$:
\begin{equation}
 C_h \in
 \operatorname*{arg\,min}_{\substack{
   E\subseteq\mathcal E(S_h)\\
   \partial E=S_h\\
   \bm\ell(E)=\bm\ell_h}}
 \sum_{e\in E}b_e.
 \label{eq:smp_edge_representation}
\end{equation}

To estimate the local posterior, we assume that the observed part $O_h$ of the motif has only two explanations: independent edge errors within the motif, or the candidate hyperedge fault. We first consider the complete four-detector motif in Fig.~\ref{fig:circuit-noise-motifs}(c), with $n_h=k_h=4$. The hyperedge alone has relative weight $P_H=p_h/(1-p_h)$. We approximate the internal independent-error explanation by one data error and one measurement error, with representative probabilities $p_{\mathrm{iid}}$ and $p_{\mathrm{meas}}$. Their odds are $P_{\mathrm{iid}}=p_{\mathrm{iid}}/(1-p_{\mathrm{iid}})$ and $P_m=p_{\mathrm{meas}}/(1-p_{\mathrm{meas}})$, giving $P_I=P_{\mathrm{iid}}P_m$. Normalizing only these two explanations gives the approximate conditional probability
\begin{equation}
 \Pr_{\mathrm{local}}(h\mid O_h=S_h)
 =\frac{P_H}{P_H+P_I}.
 \label{eq:smp_complete_local_probability}
\end{equation}
This conditional probability refers to the two-explanation model. If elementary fault probabilities scale linearly with a small noise strength $p$, then $P_H=O(p)$ and $P_I=O(p^2)$. A complete motif therefore favors the single hyperedge fault in this comparison.

For $k_h-n_h=1$, the hyperedge explanation requires an additional error to cancel the missing detection event. We estimate this contribution by assuming that the missing detector can be paired with a nearest detection event through either a measurement-error edge or a data-error edge. Each pairing represents an additional error whose detector flip cancels the hyperedge's flip at the missing detector, since detector flips add modulo two. In the $3/4$ example, we approximate the two pairing channels by independent errors with representative probabilities $p_{\mathrm{meas}}$ and $p_{\mathrm{iid}}$. We assume that each leaves the other three motif events unchanged. Exactly one of these channels must occur to cancel the missing event. Relative to neither occurring, this estimate gives
\begin{equation}
 \begin{aligned}
 P_{\mathrm{hidden}}
 &=\frac{p_{\mathrm{iid}}(1-p_{\mathrm{meas}})
       +p_{\mathrm{meas}}(1-p_{\mathrm{iid}})}
       {(1-p_{\mathrm{iid}})(1-p_{\mathrm{meas}})}\\
 &=P_{\mathrm{iid}}+P_m.
 \end{aligned}
 \label{eq:smp_hidden_weight}
\end{equation}
The hyperedge explanation then has relative weight $P_H P_{\mathrm{hidden}}$. We retain $P_I$ as a representative weight for the competing independent-error explanation, without enumerating its configurations for the incomplete motif. The resulting conditional probability is
\begin{equation}
 \Pr_{\mathrm{local}}(h\mid O_h)
 =\frac{P_H P_{\mathrm{hidden}}}
       {P_I+P_H P_{\mathrm{hidden}}},
 \qquad |O_h|=3.
 \label{eq:smp_incomplete_local_probability}
\end{equation}
Both explanations now have weight $O(p^2)$, so the incomplete motif need not strongly favor the hyperedge.

These conditional probabilities set the strength of reweighting through $\alpha_h=\eta_h/(1+\eta_h)$. For the four-detector example, the evidence ratio is
\begin{equation}
 \eta_h\approx
 \begin{cases}
  P_H/P_I, & k_h-n_h=0,\\[2pt]
  P_H P_{\mathrm{hidden}}/P_I, & k_h-n_h=1.
 \end{cases}
 \label{eq:smp_approx_motif_evidence}
\end{equation}
The approximation retains the two competing explanations and an estimated cancellation weight, while omitting the detailed connectivity and multiplicity of local edge configurations. Complete three-detector motifs use the same comparison of hyperedge and independent-error evidence. Appendix~\ref{app:willow-decoders} specifies their additional scoring convention and the missing-event score used in the benchmarks.

For an isolated cluster, stronger evidence for a single correlated fault should move the cluster cost from the sum of independent edge costs $W_{E,h}$ toward the joint fault cost $W_{C,h}$. We use the evidence ratio $\eta_h$ to weight these two descriptions. Applying the proportional cost allocation in Eq.~\eqref{eq:smp_posterior_normalized_weight} gives
\begin{equation}
 \begin{aligned}
 w_e&=(1-\alpha_h)b_e+\alpha_h\widetilde b_{h,e}\\
 &=\frac{b_e+\eta_h\widetilde b_{h,e}}{1+\eta_h},
 \end{aligned}
 \qquad e\in C_h.
 \label{eq:smp_approx_isolated}
\end{equation}
When $k_h-n_h=1$, $C_h$ still represents the complete support $S_h$. The missing-event factor controls the strength of the proposal; matching must supply any additional edges needed to cancel the detection event at the missing detector if it selects the full cluster.

\begin{figure}[!htbp]
  \centering
  \includegraphics[width=0.82\columnwidth]{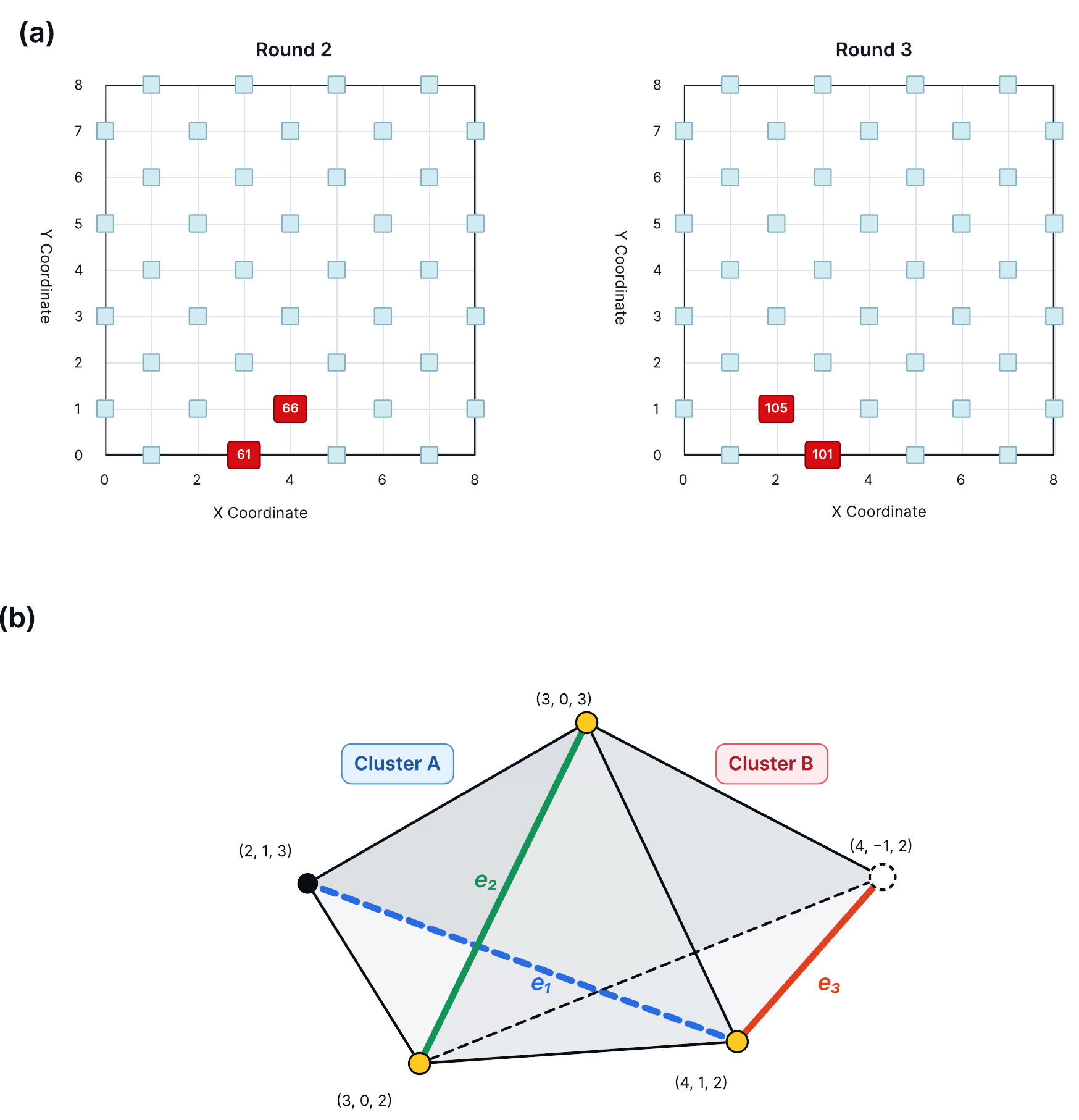}
  \caption{Overlapping clusters generated by a single correlated circuit fault. \textbf{(a)} Detector outcomes from a distance-\(5\) surface code circuit with \(r=5\) syndrome extraction rounds. The panels show rounds 2 and 3 under zero-based indexing. A correlated \(Z_3X_{12}\) fault is inserted after the second CX layer of round 2. It triggers \(D_{61}\), \(D_{66}\), \(D_{101}\), and \(D_{105}\). These events form a complete four-detector syndrome motif. \textbf{(b)} The matching graph edge sets defining the two clusters identified by SMP. Their detector graph coordinates \((2,1,3)\), \((3,0,2)\), \((3,0,3)\), and \((4,1,2)\) correspond to \(D_{105}\), \(D_{61}\), \(D_{101}\), and \(D_{66}\), respectively. Cluster A has the four-detector support \(S_A=\{(2,1,3),(3,0,2),(3,0,3),(4,1,2)\}\). Cluster B has the three-detector support \(S_B=\{(3,0,2),(3,0,3),(4,1,2)\}\). The open vertex at \((4,-1,2)\) is a boundary endpoint in the matching representation of cluster B. It is not a detector in \(S_B\). The clusters are \(C_A=\{e_1,e_2\}\) and \(C_B=\{e_2,e_3\}\). Edge \(e_2\) receives both cluster proposals.}
  \label{fig:overlap-clusters}
\end{figure}

Different syndrome motifs can be activated by the same observed events, and their clusters can share matching edges. Figure~\ref{fig:overlap-clusters}(a) shows an example in a distance-$5$ surface code. A single correlated $Z_3X_{12}$ fault triggers four detectors across two syndrome extraction rounds. These events activate both a four-detector motif and a three-detector motif contained within it. Figure~\ref{fig:overlap-clusters}(b) shows their graph representations, $C_A=\{e_1,e_2\}$ and $C_B=\{e_2,e_3\}$. The two clusters share $e_2$, so both motifs provide evidence for changing its cost.

We seek an edge cost that reflects the evidence from all contributing motifs. For an isolated cluster, Eq.~\eqref{eq:smp_approx_isolated} averages the original cost and the cluster's proposed cost with relative weights $1$ and $\eta_h$. We extend this prescription to a shared edge by assigning each proposal its own evidence ratio $\eta_h$ and retaining the original cost with unit relative weight. Thus, stronger evidence gives a proposal a larger share, while weak evidence leaves the cost near its original value. Let $\mathcal H_e=\{h:e\in C_h\}$ index the activated clusters containing edge $e$. Normalizing these relative weights gives
\begin{equation}
 w_e=\frac{b_e+\sum_{h\in\mathcal H_e}
                  \eta_h\widetilde b_{h,e}}
              {1+\sum_{h\in\mathcal H_e}\eta_h}.
 \label{eq:smp_overlap_weight}
\end{equation}
With one contributing cluster, this expression reduces to Eq.~\eqref{eq:smp_approx_isolated}; with none, $w_e=b_e$.

For the two clusters in Fig.~\ref{fig:overlap-clusters}(b), the rule gives
\begin{equation}
 \begin{aligned}
  w_{e_1}
   &=\frac{b_{e_1}+\eta_A\widetilde b_{A,e_1}}
           {1+\eta_A},\\[5pt]
  w_{e_2}
   &=\frac{b_{e_2}+\eta_A\widetilde b_{A,e_2}
                    +\eta_B\widetilde b_{B,e_2}}
           {1+\eta_A+\eta_B},\\[5pt]
  w_{e_3}
   &=\frac{b_{e_3}+\eta_B\widetilde b_{B,e_3}}
           {1+\eta_B}.
 \end{aligned}
 \label{eq:smp_overlap_example}
\end{equation}
The shared edge $e_2$ combines evidence from both clusters, while $e_1$ and $e_3$ each receive one proposal. This rule approximates the regional normalization in Appendix~\ref{app:local-posterior} by mixing costs separately on each edge. It does not construct a joint posterior over compatible combinations of complete clusters.

\subsection{Decoder workflow and complexity analysis}
\label{sec:workflow}

For the native surface-code setting, SMP is implemented in the following four steps.

\begin{enumerate}
  \item Select the three- and four-detector motifs $S_h$ defined in Sec.~\ref{sec:smp_posterior}, with candidate priors $p_h$ obtained from the DEM as specified in Appendix~\ref{app:willow-decoders}.

  \item Construct each cluster $C_h$ using Eq.~\eqref{eq:smp_edge_representation} to reproduce the candidate's detector support $S_h$ and logical effect $\bm\ell_h$. Obtain its cost proposal from Eqs.~\eqref{eq:smp_posterior_cost_reference} and \eqref{eq:smp_posterior_normalized_weight}, retaining only valid clusters with $0\leq W_{C,h}<W_{E,h}$.

  \item For the measured syndrome, evaluate the motif evidence using Eq.~\eqref{eq:smp_approx_motif_evidence} and the three-detector rule in Appendix~\ref{app:willow-decoders}. Update isolated clusters using Eq.~\eqref{eq:smp_approx_isolated} and combine overlapping proposals using Eq.~\eqref{eq:smp_overlap_weight}.

  \item Decode the complete syndrome on the reweighted graph using the chosen matching decoder, such as standard or correlated MWPM, to infer the logical-observable flips.
\end{enumerate}

For a fixed DEM, each retained cluster $C_h$ and its proposals $\widetilde b_{h,e}$ are constructed once. For each syndrome $Y$, evaluating $\eta_h(Y)$ and accumulating its contributions to Eq.~\eqref{eq:smp_overlap_weight} takes $O(k_h+|C_h|)$ work per candidate. In the surface-code circuits considered here, $k_h=|S_h|$, $|C_h|$, and the number of candidates containing any detector are bounded independently of $d$ and $r$. Thus, $|\mathcal H|=O(|\mathcal V_{\mathrm{det}}|)=O(d^2r)$. SMP therefore adds $O(d^2r)$ preprocessing time per syndrome and $O(d^2r)$ storage for the motifs, clusters, and proposals. The matching step retains the complexity of the chosen backend.

\subsection{Applying SMP to different codes and circuits}
\label{sec:smp_general}

Extending SMP beyond the native surface-code setting requires two conditions on the fault representation. First, a candidate fault $h$ must have a known detector support $S_h$, so that the observed part $O_h=S_h\cap Y$ of its motif can provide local evidence $\eta_h$. Second, the fault must admit an edge representation in the chosen matching graph or graphs. This representation must preserve its detector support under the corresponding projection and induce the same logical-observable flips, as specified by Eq.~\eqref{eq:smp_general_representation}.

Two nontrivial situations arise when transferring this physical evidence to matching graphs:
\begin{itemize}
  \item The detector support $S_h$ of one physical fault spans several matching graphs, so no single graph contains the complete motif.
  \item The graph structure changes under the matching representation, so the detector boundary of $C_h$ represents a mapped version of $S_h$ rather than the original detector set itself.
\end{itemize}
These two situations can occur together: a fault's detector support can be mapped to a different representation and distributed across several matching graphs. The M\"obius and logical observable matching applications below illustrate how SMP transfers physical motif evidence to the corresponding graph representations. We first describe how to preserve the fault's detector and logical effects under projection, then explain how to share its evidence across graphs.

\paragraph{Different matching representations.}
The physical fault and its syndrome motif do not depend on the matching representation, but their graph representation can. Index the matching graphs by $j$. Let $G_j$ have edge set $\mathcal E_j$, and let $\iota_j(S_h)$ denote the detector support of $h$ in that graph. Here $\partial$ selects the graph's detector vertices incident to an odd number of selected edges and excludes boundary vertices. A valid cluster $C_h^{(j)}\subseteq\mathcal E_j$ must satisfy
\begin{equation}
 \partial C_h^{(j)}=\iota_j(S_h),
 \qquad
 \bm\ell_j(C_h^{(j)})=\bm\ell_{h,j}.
 \label{eq:smp_general_representation}
\end{equation}
Here $\bm\ell_{h,j}$ is the logical-observable flip vector of $h$ for the observables tracked by graph $j$. The map $\bm\ell_j$ gives the logical-observable flip vector reconstructed from the selected edges by that backend, including any required lifting. The two conditions preserve the projected detector support $\iota_j(S_h)$ and logical effect $\bm\ell_{h,j}$.

The construction of $C_h^{(j)}$ is therefore backend dependent. For the surface code, the cluster is constructed directly in the native matching graph using Eq.~\eqref{eq:smp_edge_representation}. For the color code, the same principle is applied in the M\"obius representation. If no parity- and logical-equivalent cluster exists, that candidate fault cannot be incorporated into SMP through the chosen matching representation.

\paragraph{Motifs spanning several matching graphs.}
A physical fault can affect several matching problems, as in logical-gate circuits. We evaluate its local evidence $\eta_h(Y)$ once from the physical syndrome $Y$, before any projection. We compare local explanations with and without $h$ using Eq.~\eqref{eq:smp_local_posterior_share}. The approximation for these explanations is specified in each application. Every graph receiving the fault uses the same evidence ratio.

In graph $G_j$, the candidate has a cluster $C_h^{(j)}$ satisfying Eq.~\eqref{eq:smp_general_representation}. Its original edge weights $b_e^{(j)}$ give the cluster cost $W_{E,h}^{(j)}$, as in Eq.~\eqref{eq:smp_posterior_cost_reference}. The proposed cluster cost change depends on the matching representation, as specified below for the color code and LOM. We distribute this change among the cluster edges in proportion to their original weights, following the allocation principle of Eq.~\eqref{eq:smp_posterior_normalized_weight}. The shared evidence controls the interpolation between original and proposed edge costs. Overlapping proposals are combined using Eq.~\eqref{eq:smp_overlap_weight}, and each graph is decoded with its updated weights.

\begin{figure*}[tp]
\centering
\includegraphics[width=0.90\textwidth]{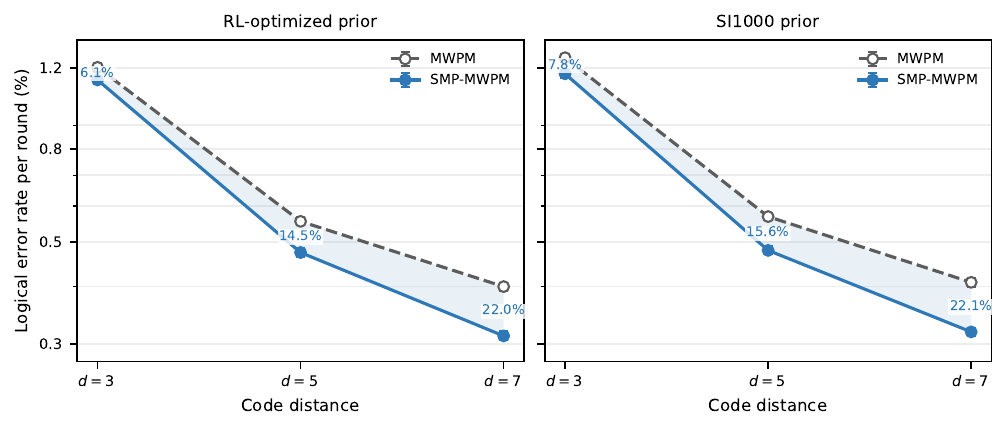}
\caption{Logical error rates per round for MWPM and SMP-MWPM on the Willow data. The panels use the reinforcement-learning-optimized (RL) prior and the SI1000 noise model prior. At each distance, the logical failure probability per memory experiment \(P_{\mathrm L}(r)\) is measured after \(r=10,13,30,50\) extraction rounds, and the per-round rate \(\epsilon\) is extracted from Eq.~\eqref{eq:per-round-fit}. Logical-\(X\) and logical-\(Z\) data are fitted separately and the resulting rates are averaged. Error bars show 95\% intervals obtained from 5,000 parametric bootstrap resamples. Shading marks the gap between the two curves, and labels give the relative reduction \(R=1-\epsilon_{\mathrm{SMP}}/\epsilon_{\mathrm{base}}\), expressed as a percentage. Each input round count contains 50,000 shots per logical basis.}
\label{fig:willow-mwpm}
\end{figure*}

\begin{figure*}[tp]
\centering
\includegraphics[width=0.90\textwidth]{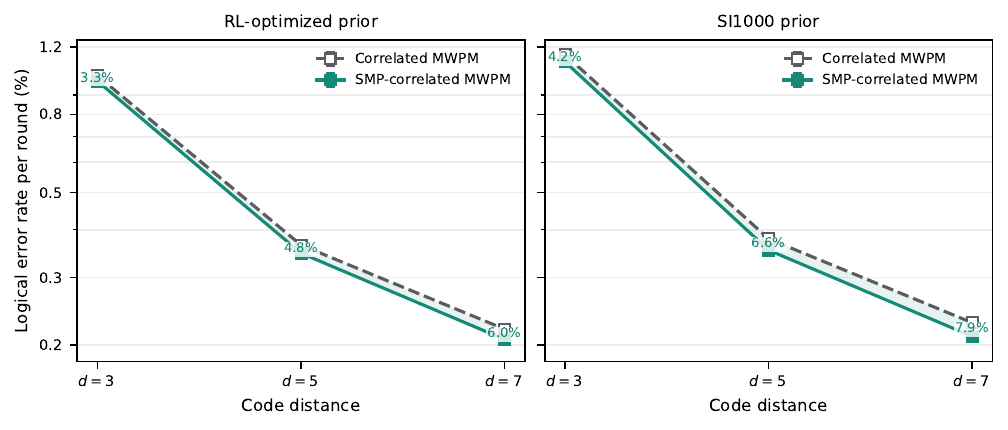}
\caption{Logical error rates per round for correlated MWPM and SMP-correlated MWPM on the Willow data. The panels use the reinforcement-learning-optimized (RL) prior and the SI1000 noise model prior with input round counts $r=10,13,30,50$. The per-round rate \(\epsilon\) is extracted independently for each distance, prior, logical basis, and decoder using Eq.~\eqref{eq:per-round-fit}. Logical-\(X\) and logical-\(Z\) rates are averaged after fitting. Error bars show 95\% intervals obtained from 5,000 parametric bootstrap resamples. Shading marks the gap between the two curves, and labels give the relative reduction \(R=1-\epsilon_{\mathrm{SMP}}/\epsilon_{\mathrm{base}}\), expressed as a percentage. Each input round count contains 50,000 shots per logical basis.}
\label{fig:willow-correlated}
\end{figure*}

\section{Surface code memory decoding}
\label{sec:surface}
In the Willow evaluations, we directly reuse the publicly released DEM and learned priors, without any retraining \cite{willow2025}. The dataset contains surface code memory experiments at distances \(d=3,5,\) and \(7\), measured in both the logical-\(X\) and logical-\(Z\) bases. We use circuits with \(r=10,13,30,\) and \(50\) syndrome extraction rounds, with 50,000 shots for each logical basis. For each dataset, we decode the syndrome using either the SI1000 prior or the reinforcement-learning-optimized (RL) prior provided with the public data.

To compare experiments with different numbers of rounds, we extract the logical error rate per round, \(\epsilon\), from the decay of the logical fidelity \cite{experiment_quantum_ai_suppressing_2023,willow2025,bausch2024learning}. For the logical failure probability per memory experiment, \(P_{\mathrm L}(r)\), with \(r\) rounds, we fit
\begin{equation}
F(r)=1-2P_{\mathrm L}(r)
=F_0(1-2\epsilon)^r .
\label{eq:per-round-fit}
\end{equation}
Here $F_0$ is a fitted prefactor. The two logical bases are fitted separately and their rates are averaged. Appendix~\ref{app:willow-decoders} gives the fitting and decoder settings. Figures~\ref{fig:willow-mwpm} and \ref{fig:willow-correlated} use the fitted logical error rate per round \(\epsilon\) to summarize decoder performance across the measured round counts, \(r=10,13,30,50\).

SMP improves MWPM at every tested code distance under both priors (Fig.~\ref{fig:willow-mwpm}). We quantify this improvement by the relative reduction
\begin{equation}
R=
\frac{\epsilon_{\mathrm{base}}-\epsilon_{\mathrm{SMP}}}
{\epsilon_{\mathrm{base}}}.
\label{eq:relative-reduction}
\end{equation}
Here $\epsilon_{\mathrm{base}}$ and $\epsilon_{\mathrm{SMP}}$ are the logical error rates per round of the baseline decoder and its SMP counterpart. For the comparisons below, the same definition uses the logical failure probability per memory experiment $P_{\mathrm L}$ or the circuit failure probability $P_{\mathrm{fail}}$ in place of $\epsilon$. The quantity $R$ is dimensionless; for example, $R=0.24$ corresponds to a 24\% reduction. Relative reductions are expressed as percentages throughout the figures and tables.

 The gain grows with code distance: under the RL (SI1000) prior, $R$ increases from $6.1\%$ ($7.8\%$) at $d=3$ to $22.0\%$ ($22.1\%$) at $d=7$.

SMP also improves correlated MWPM, even though this decoder already uses reweighting to recover part of the error correlations (Fig.~\ref{fig:willow-correlated}). The improvement persists at all three code distances under both priors. At $d=7$, SMP further reduces the logical error rate per round by $6.0\%$ with the RL prior and $7.9\%$ with SI1000. This additional gain indicates that SMP extracts useful correlation information that correlated MWPM does not fully capture during its own reweighting procedure.

The broader comparison in Table~\ref{tab:willow-selected} further demonstrates the accuracy of SMP-correlated MWPM. On the $d=7$, $r=13$ Willow data, it achieves lower observed logical failure probabilities than both Google's released correlated matching and belief matching with at most 10 BP iterations, under both priors. Among the entries shown, the Harmony-51 ensemble has the lowest observed failure probability. Unlike the curves, this table reports the probability of failure over the entire 13-round experiment. Decoder settings are given in Appendix~\ref{app:willow-decoders}.

\begin{table}[!htbp]
\caption{Decoder comparison on the $d=7$, $r=13$ Willow data. Entries are pooled logical failure probabilities in percent, combining 50,000 shots from each logical basis. All methods use the prior indicated by the column. These probabilities refer to entire memory experiments, rather than the fitted per-round rates in Figs.~\ref{fig:willow-mwpm} and \ref{fig:willow-correlated}. Decoder definitions are given in Appendix~\ref{app:willow-decoders}.}
\label{tab:willow-selected}
\centering\small
\begin{tabular*}{\columnwidth}{@{\extracolsep{\fill}}lrr@{}}
\toprule
Decoder & SI1000 & RL \\
\midrule
\shortstack[l]{Correlated matching\\(Google released)} & 2.415 & 2.274 \\
Harmony-51 & 2.026 & 1.916 \\
Belief matching (BP: max. 10) & 2.541 & 2.215 \\
Correlated MWPM & 2.487 & 2.379 \\
SMP-correlated MWPM & \textbf{2.265} & \textbf{2.171} \\
\bottomrule
\end{tabular*}
\end{table}

Further, we use Stim to simulate unrotated surface code memory circuits in the logical-$Z$ basis and test whether the advantage persists beyond the Willow code distances \cite{gidney2021stim}. We apply uniform circuit-level depolarizing noise with strength $p$. Each CNOT is followed by the two-qubit depolarizing channel in Eq.~\eqref{eq:circuit-depolarizing-channel} of Sec.~\ref{sec:background}, so each of the 15 nonidentity two-qubit Pauli errors occurs with probability $p/15$. Single-qubit depolarizing noise is applied after each single-qubit Clifford gate and to each data qubit before every syndrome extraction round, with $X$, $Y$, and $Z$ errors each occurring with probability $p/3$. Measurement errors and reset errors each occur with probability $p$. We set $r=d$ and supply the same 20,000 sampled shots at each simulation point to the decoders being compared.

\begin{figure}[!htbp]
  \centering
  \includegraphics[width=\columnwidth]{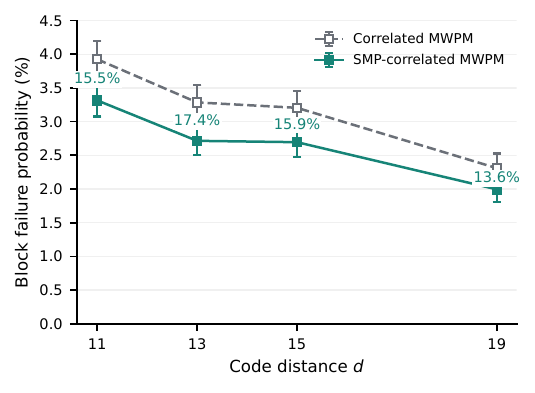}
  \caption{Logical failure probabilities per memory experiment at $p=0.0081578947$ (0.8158\%) under uniform circuit-level depolarizing noise. The circuit is unrotated logical-$Z$ memory with $d=11,13,15,19$ and $r=d$. Both decoders use the same 20,000 shots at each distance. Open gray squares with dashed lines show correlated MWPM. Filled teal squares with solid lines show SMP-correlated MWPM. Points show observed logical failure probabilities per memory experiment; vertical bars are marginal Wilson 95\% intervals. Lines guide the eye. Labels give the relative reduction from correlated MWPM to SMP-correlated MWPM. SMP gives the lower point estimate at all four distances, with reductions of 13.6--17.4\%.}
  \label{fig:depolarizing}
\end{figure}

In the four-decoder scan, SMP lowers the logical failure probability at all points with at least 100 failures in the corresponding baseline: 44 points for MWPM and 37 for correlated MWPM. We then extend the correlated MWPM comparison to $d=11,13,15,19$ over $p=0.005$--$0.01$. Figure~\ref{fig:depolarizing} shows the results at $p\simeq0.00816$, where SMP reduces the logical failure probability by $13.6$--$17.4\%$ across these four distances. The improvement through $d=19$ shows that the benefit of syndrome-based correlation information persists beyond the sizes available in the Willow experiments. Appendix~\ref{app:surface-scan} shows the noise scan over $0.0065\leq p\leq0.01$; the complete scan remains in the accompanying comma-separated values (CSV) file.

\section{Color code memory decoding}
\label{sec:color}

\subsection{The color code decoding problem}
\label{sec:color-problem}
Two-dimensional color codes combine local stabilizer checks with transversal Clifford gates, offering a route to low-overhead logical computation \cite{bombin2006color,fowler2011color,landahl2011color,thomsen2024color,beverland2021cost}. Their experimental promise is demonstrated by fault-tolerant gate operations and syndrome extraction in trapped ions, and by memory scaling and logical operations in superconducting processors \cite{postler2022universal,postler2024steane,lacroix2025color}.

\begin{figure}[tb]
\centering
\includegraphics[width=0.85\columnwidth]{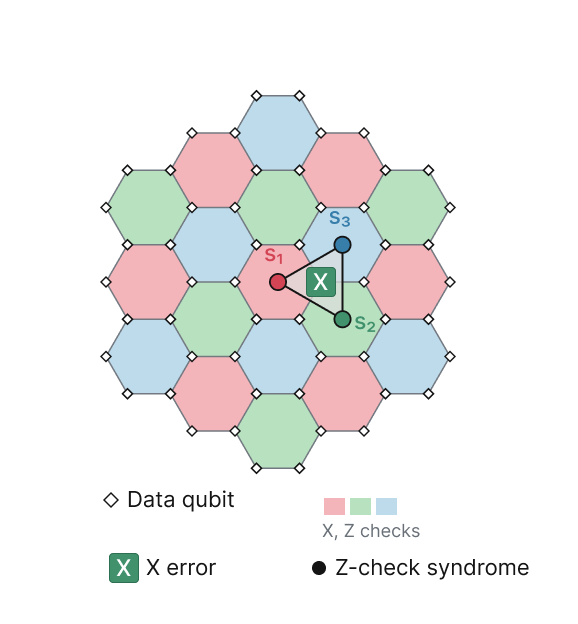}
\caption{Structure of the $6.6.6$ color code and the syndrome produced by a single data qubit error. Diamonds denote data qubits. Red, green, and blue faces label color sectors. Each face supports both $X$- and $Z$-type stabilizers on its six vertices. With ideal measurements and no other errors, the marked $X$ error flips the three $Z$ checks that act on that qubit and leaves the $X$ checks unchanged. A $Z$ error at the same location instead flips the three neighboring $X$ checks and leaves the $Z$ checks unchanged. Filled circles $s_1$, $s_2$, and $s_3$ mark the flipped checks for the illustrated $X$ error. The gray triangle denotes their common fault support $S_h$, not three independent matching edges. The outer contour bounds the illustrated bulk region; it is not a code boundary.}
\label{fig:color666-syndrome}
\end{figure}

Decoding these experiments requires accounting for correlations between checks of different colors. In the $6.6.6$ color code shown in Fig.~\ref{fig:color666-syndrome}, data qubits lie at vertices, and each colored face supports both $X$- and $Z$-type stabilizers. An $X$ error on a bulk qubit therefore flips the three neighboring $Z$ checks, one of each color, producing the motif $S_h=\{s_1,s_2,s_3\}$. A $Z$ error at the same location produces the corresponding motif in the three neighboring $X$ checks. Thus, even with ideal measurements, one fault correlates three syndrome events. Propagation through the extraction circuit can extend this correlation to larger detector motifs \cite{gidney2023chromobius,lee2025color}. Representing these faults in a matching graph requires mapping their multi-detector supports to edges with at most two detector endpoints.

M\"obius matching provides such a representation by joining subgraphs for pairs of check colors at their boundaries \cite{sahay2022mobius}. Each subgraph contains detectors from two of the three colors. Joining these subgraphs allows matching to connect syndrome information across color sectors. Chromobius implements this construction for circuit-level noise using efficient minimum-weight matching \cite{gidney2023chromobius}. It thus combines a graph representation of the multi-color syndrome with a decoder that accounts for faults in the extraction circuit. We therefore use Chromobius as the decoding backend and apply SMP to its edge weights. The updated graph is decoded through its matching and lifting procedures.

We use the native CX implementation of the Middle-out extraction circuit. This circuit requires no dedicated ancilla qubits and uses a hexagonal layout with three neighbors per qubit \cite{gidney2023chromobius}. It extracts checks by folding stabilizers into single-qubit operators for measurement and then unfolding them. Alternating cycles extract complementary checks. With Chromobius decoding, this circuit outperformed the other color code circuits compared by Gidney and Jones at $p=0.001$, despite a reduced circuit distance \cite{gidney2023chromobius}.

We map the faults in this circuit to the native M\"obius graph according to Eq.~\eqref{eq:smp_general_representation}. Each physical detector appears in two pair-of-color sectors. These copies define the map $\iota(S)$, which contains both copies of every detector in $S$. For each fault $h$, we construct a graph cluster $C_h$ satisfying $\partial C_h=\iota(S_h)$ and $\bm\ell(C_h)=\bm\ell_h$. The first condition reproduces the duplicated detector support. The second preserves the fault's logical-observable flips, with $\bm\ell(C_h)$ obtained through the original lifting procedure.

Detector copies represent the same physical outcomes, so we evaluate motif evidence before duplication. We approximate the contribution of other faults by independent background flips with a common probability. We obtain this probability by averaging the detector marginals after removing candidate $h$ from the DEM. This \emph{uniform-hidden} approximation uses the observed count $n_h=|O_h|$ to compare the independent-error and hyperedge explanations. The common probability averages over spatial variation, and the independence assumption neglects background correlations. Appendix~\ref{app:color} gives the resulting evidence ratio $\eta_h(Y)$ and activation thresholds.

\begin{figure*}[tp]
\centering
\includegraphics[width=\textwidth]{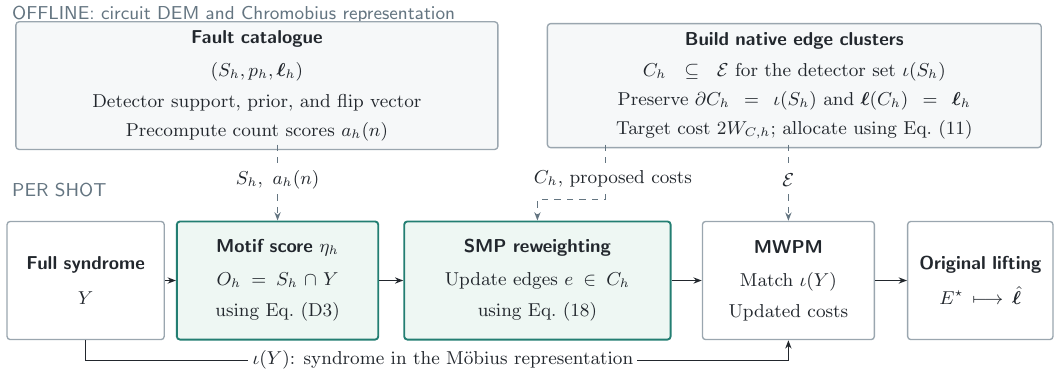}
\caption{SMP with Chromobius. Evidence is evaluated on the observed part $O_h=S_h\cap Y$ of the physical motif before each detector is duplicated in the M\"obius graph. The cluster $C_h$ represents $\iota(S_h)$ and receives target cost $2W_{C,h}$. Motif evidence controls the edge reweighting; matching and lifting give the logical-observable flip prediction. Teal boxes mark SMP; dashed arrows carry precomputed information.}
\label{fig:color-workflow}
\end{figure*}

For each candidate, SMP assigns a target cost to its graph cluster and uses the motif evidence to control the weight update. Under the doubled-cost convention used here, this target is $2W_{C,h}$. We distribute the target cost over $C_h$ using Eq.~\eqref{eq:smp_posterior_normalized_weight}. We then combine proposals from overlapping clusters using their motif evidence in Eq.~\eqref{eq:smp_overlap_weight}. Chromobius matches $\iota(Y)$ with the updated weights and lifts the result to the predicted logical-observable flip vector $\hat{\bm\ell}$ [Fig.~\ref{fig:color-workflow}].

\subsection{Numerical results}
\label{sec:color-results}
We simulate Middle-out color code memory circuits at $d=9,11,13$ in both logical bases, with $r=2d$ extraction rounds. Here $d$ specifies the base width of the triangular patch; the Middle-out circuit has a smaller circuit-level distance. We compare SMP with official Chromobius under uniform circuit-level depolarizing noise at $p=0.001$ and $p=0.003$. Both decoders use the same circuit, prior, and 100,000 shots at each point. Appendix~\ref{app:color} records the sample provenance and projection checks.

\begin{table}[!htbp]
\caption{Relative reduction $R$ (percent) from SMP in Middle-out color code memory. Both noise strengths use the current uniform-hidden implementation, $r=2d$, and 100,000 shared shots per basis and patch size. Logical failure probabilities and paired statistics are given in Table~\ref{tab:color}.}
\label{tab:color-noise}
\centering
\begin{ruledtabular}
\begin{tabular}{crrrr}
& \multicolumn{2}{c}{$p=0.001$} & \multicolumn{2}{c}{$p=0.003$} \\
$d$ & $X$ & $Z$ & $X$ & $Z$ \\
\hline
\ColorReductionRows
\end{tabular}
\end{ruledtabular}
\end{table}

\begin{figure*}[tp]
\centering
\includegraphics[width=\textwidth]{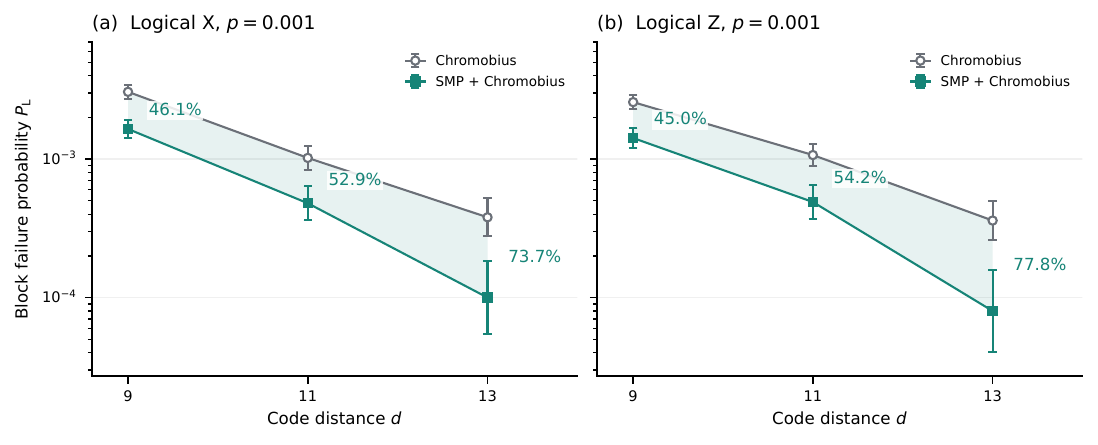}
\caption{Color code memory decoding under circuit-level depolarizing noise. We use the Middle-out syndrome extraction circuit \cite{gidney2023chromobius}. \textbf{(a,b)} Logical failure probabilities per memory experiment for logical $X$ and $Z$ at $p=0.001$, with the same vertical scale. Each point uses $r=2d$ and 100,000 shared shots. Open circles show official Chromobius; filled squares show uniform-hidden SMP followed by the same matching and lifting. Error bars are marginal Wilson 95\% intervals. Shading marks the gap between the curves. Labels give the relative reduction $R=1-P_{\mathrm{L,SMP}}/P_{\mathrm{L,base}}$, expressed as a percentage. Table~\ref{tab:color-noise} compares the reductions at two noise strengths.}
\label{fig:color}
\end{figure*}

SMP lowers the observed failure probability at every tested configuration. At $p=0.001$, the relative reduction grows from about 45--46\% at $d=9$ to 74--78\% at $d=13$ in the two logical bases [Fig.~\ref{fig:color}]. At $d=13$, the logical-$Z$ failure probability falls from 0.036\% to 0.008\%. These estimates are based on few failures, with uncertainty shown by the Wilson intervals in the figure [full results and paired statistics in Table~\ref{tab:color}]. Together, these results support using physical syndrome motifs to update matching weights in the M\"obius representation.

At $p=0.003$, SMP reduces the failure probability by about 29\% at $d=9$ and 43--44\% at $d=13$ [Table~\ref{tab:color-noise}]. These relative gains are smaller than those at $p=0.001$. This trend is consistent with the local comparison in Sec.~\ref{sec:smp_approximation}, where lower noise favors a complete single-fault explanation over competing multiple-fault explanations.

\section{Decoding fast transversal logical gates}
\label{sec:gates}

\subsection{Decoding across transversal logical gates}
\label{sec:gates-problem}
Transversal gates offer a route to fast logical operations when hardware supports interactions between code blocks, as demonstrated in reconfigurable atom arrays \cite{bluvstein2024logical}. Realizing this speed advantage requires reducing the number of error correction rounds between gates. Joint decoding across code blocks can reduce this overhead by accounting for fault correlations across the gates \cite{cain2024correlated}. This principle underlies transversal algorithmic fault tolerance, which supports logical computation with a constant number of extraction rounds between operations \cite{zhou2025transversal}. To decode these correlations, matching-based approaches track error propagation between blocks \cite{wan2024iterative,sahay2025transversal,cain2025fastcorrelated}. Among these approaches, we use logical observable matching (LOM) for its high decoding accuracy in transversal Clifford benchmarks and efficient matching for each logical measurement \cite{serraperalta2026lom}.

A transversal CNOT creates cross-block syndrome correlations through both error propagation and the detector parity relations. Consider a control patch A and a target patch B, with physical CNOTs coupling corresponding data qubits. The propagating errors obey
\begin{equation}
 X_A\mapsto X_AX_B,\qquad Z_B\mapsto Z_AZ_B,
 \label{eq:cnot-propagation}
\end{equation}
while $Z_A$ and $X_B$ remain on their original patch. The detector parities also depend on how the gate transforms the measured stabilizers. These parity constraints are formalized by circuit-to-code constructions and spacetime codes \cite{bacon2017sparse,delfosse2023spacetime,li2025ldpccircuits}. For corresponding $Z$ stabilizers, the transformation is $S_Z^A\mapsto S_Z^A$ and $S_Z^B\mapsto S_Z^AS_Z^B$. Let $s_A^\pm$ and $s_B^\pm$ be their binary outcomes before ($-$) and after ($+$) the gate. With no faults, $s_A^+=s_A^-$ and $s_B^+=s_A^-\oplus s_B^-$. LOM uses a pre-gate detector frame, in which the target patch parity is \cite{serraperalta2026lom}
\[
 d_B=s_A^+\oplus s_B^+\oplus s_B^-=0.
\]
This detector assigned to B contains the outcome $s_A^+$ from A. A measurement error on A can therefore trigger a detector on B through their shared measurement record.

\begin{figure*}[tp]
\centering
\includegraphics[width=\textwidth]{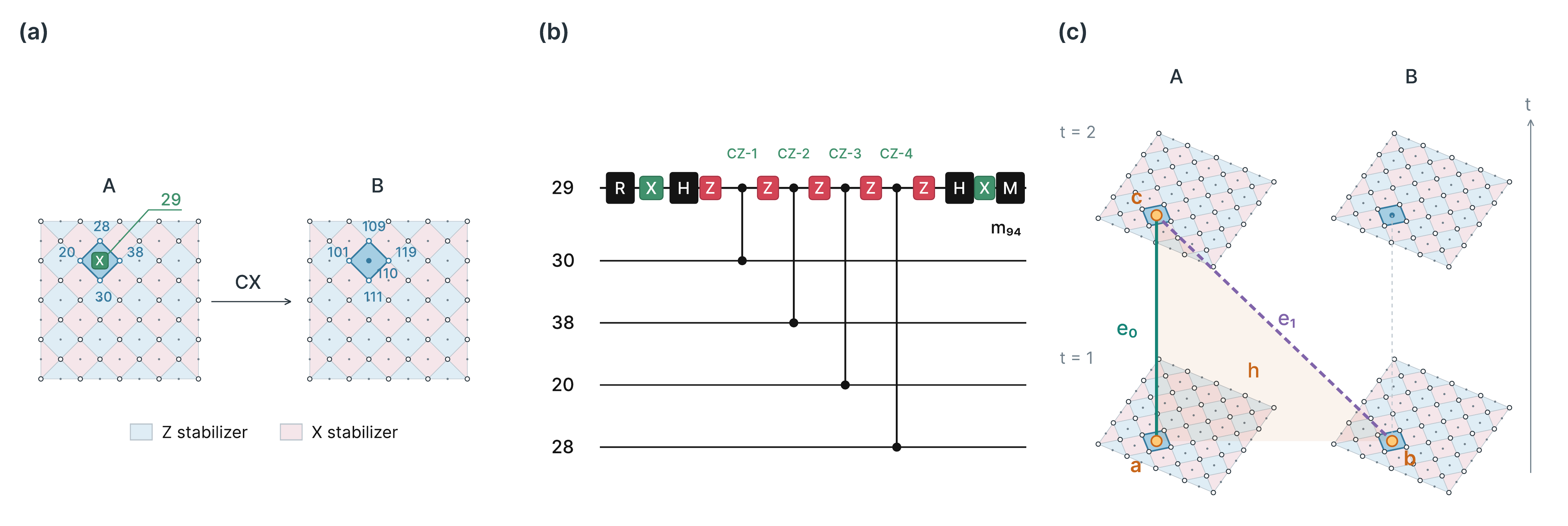}
\caption{Fault evolution and syndrome formation in a transversal CNOT circuit, with the corresponding edge sets in logical observable matching (LOM).
\textbf{(a)} Unrotated $d=5$ patches A and B are the control and target of a transversal CNOT.
Open circles denote data qubits; small grey dots denote check ancillas.
Blue and pink faces show $Z$- and $X$-stabilizer supports.
Ancilla 29 measures $Z_{20}Z_{28}Z_{30}Z_{38}$ on A.
The green $X$ marks a fault immediately after its reset in the QEC round following the CNOT.
\textbf{(b)} Gates acting on ancilla 29 are shown in time order.
$R$, $H$, and $M$ denote reset to $\lvert0\rangle$, Hadamard, and $Z$-basis measurement.
All four CZ gates couple qubits within A, in the order 30, 38, 20, 28.
Idles and gates acting only on data qubits are omitted.
The green $X$ and red $Z$ markers track the same fault.
The first $H$ maps $X$ to $Z$, which commutes through the CZ gates.
The final $H$ maps it back to $X$, flipping only record $m_{94}$ and producing no data qubit error.
\textbf{(c)} The events are $a=D_{94}$ at $(3,2,1)$, $b=D_{134}$ at $(3,12,1)$, and $c=D_{174}$ at $(3,2,2)$.
Their parities are $m_{94}\oplus m_{14}$, $m_{94}\oplus m_{134}\oplus m_{54}$, and $m_{174}\oplus m_{94}$, respectively.
The shared record $m_{94}$ flips all three.
The orange membrane marks the physical detector support $S_h=\{a,b,c\}$ of the ancilla fault.
The graphs $G_0$ and $G_1$ correspond to the two target logical observables. The corresponding LOM edge sets are $C_h^{(0)}=\{e_0\}$ in $G_0$ and $C_h^{(1)}=\{e_1\}$ in $G_1$. The teal edge $e_0=\{a,c\}$ and dashed purple edge $e_1=\{b,c\}$ reproduce the projected detector supports in these graphs.
The edges are overlaid to locate their endpoints; the two graphs are matched independently.
Qubit, measurement, and detector indices are zero-based.}
\label{fig:cnot-motif}
\end{figure*}

Figure~\ref{fig:cnot-motif} illustrates how this shared record produces a hyperedge across code blocks. A fault on a $Z$-check ancilla after the CNOT remains on the ancilla and flips one measurement record. That record enters three detector parities, so the fault produces three events across two patches and adjacent times. The same mechanism appears in Fig.~2 of Ref.~\cite{serraperalta2026lom} and Fig.~2(b) of Ref.~\cite{cain2025fastcorrelated}. Splitting these events into pairwise edges can lose their common fault cost or connectivity \cite{serraperalta2026lom,delfosse2023splitting}.

LOM represents these faults by constructing a matching graph for each target logical measurement $O$. It first propagates $O$ backward through the circuit to identify the relevant patches and check types at each time. These checks define a detector set $V_O$, and each fault support is restricted to this set through the map $\iota_O(S)=S\cap V_O$. The graph index in Sec.~\ref{sec:smp_general} is therefore $j=O$. Under the basic model of independent data-$X$/$Z$ faults and measurement errors, each projected support has at most two endpoints and can be represented by an edge \cite{serraperalta2026lom}. For circuit-level noise, the LOM implementation supplies a decomposition into matching edges. The resulting graph $G_O$ has edge set $\mathcal E_O$, including boundary edges. On this graph, matching selects a minimum-weight edge set $E_O^\star$ with $\partial E_O^\star=Y\cap V_O$ and predicts the logical-observable flip $\hat\ell_O$. Repeating this procedure for the independent logical measurements gives the circuit prediction.

This restriction can discard evidence that identifies a fault. In Fig.~\ref{fig:cnot-motif}(c), the complete support is $S_h=\{a,b,c\}$. The graph $G_0$ for observable $O_0$ retains the edge $e_0=\{a,c\}$ and omits $b$. The graph $G_1$ for $O_1$ retains $e_1=\{b,c\}$ and omits $a$. Each projected contribution is absorbed into one matching edge, so each graph contains only part of the three-event pattern. Applying correlated MWPM independently within these graphs still limits its updates to the detectors and correlations retained in each graph. The complete motif therefore supplies additional evidence for their common fault origin.

SMP uses this additional evidence by evaluating $\eta_h(Y)$ on the observed part $O_h=S_h\cap Y$ of the physical motif before restricting the syndrome. We compute the evidence using the motif and missing-event estimates of Sec.~\ref{sec:smp_approximation}. These estimates refer to the complete physical support $S_h$, with missing events identified by $S_h\setminus Y$. The resulting evidence is shared by all graphs receiving the fault. To use it in $G_O$, we retain a native cluster $C_h^{\mathrm{LOM}}$ that satisfies Eq.~\eqref{eq:smp_general_representation} with $j=O$. This cluster reproduces $S_h\cap V_O$ and preserves the logical-observable flip $\ell_{h,O}$ of the target observable $O$. Its edges receive the weight updates determined by the shared evidence.

To set the proposed cost reduction on these edges, we compare costs for the complete fault support. The prior fault cost $W_{C,h}$ in Eq.~\eqref{eq:smp_posterior_cost_reference} describes the complete fault, whereas the native cluster cost $W_{E,h}^{(O)}$ describes only its projection. We therefore compare the prior with the cost $W_{\mathrm{full},h}$ of matching $S_h$ in the standard matching graph obtained by splitting hyperedges. The proposed cost change is
\begin{equation}
 \delta_h=W_{C,h}-W_{\mathrm{full},h}.
 \label{eq:lom-discount}
\end{equation}
The reference matching minimizes the cost over unrestricted paths without constraining the logical-observable flip vector. This minimum sets the discount applied to the native cluster, whose edges preserve the projected detector support $S_h\cap V_O$ and logical effect $\ell_{h,O}$. For finite negative $\delta_h$ and positive $W_{E,h}^{(O)}$, we distribute the cost change among these edges in proportion to their original weights, following Sec.~\ref{sec:smp_general}. A single-edge projection receives the full reduction.

\begin{figure*}[tp]
\centering
\includegraphics[width=\textwidth]{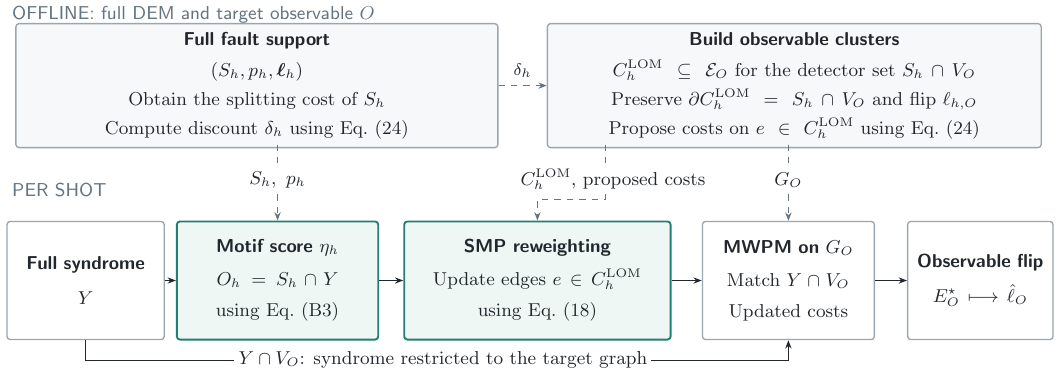}
\caption{SMP in LOM for one observable $O$. Full-motif evidence uses $O_h=S_h\cap Y$ and Eq.~\eqref{eq:surface-implementation-score}. Each cluster $C_h^{\mathrm{LOM}}\subseteq\mathcal E_O$ represents $S_h\cap V_O$ and preserves the target logical-observable flip $\ell_{h,O}$. Equation~\eqref{eq:lom-discount} proposes costs on its edges. Equation~\eqref{eq:smp_overlap_weight} combines activated proposals from overlapping clusters. Matching receives the restricted syndrome $Y\cap V_O$ and the updated costs. Each observable is decoded independently. Teal boxes mark SMP; dashed arrows carry precomputed information.}
\label{fig:lom-workflow}
\end{figure*}

We then combine proposals from overlapping clusters using their evidence $\eta_h(Y)$ in Eq.~\eqref{eq:smp_overlap_weight}. Each graph is decoded with the updated weights and its restricted syndrome $Y\cap V_O$, completing the workflow in Fig.~\ref{fig:lom-workflow}. Appendix~\ref{app:gates} gives the scoring parameters, implementation safeguards, and verification of the projected example in Fig.~\ref{fig:cnot-motif}.

\subsection{Numerical results}
\label{sec:gates-results}
\begin{figure}[!htbp]
\centering
\includegraphics[width=\columnwidth]{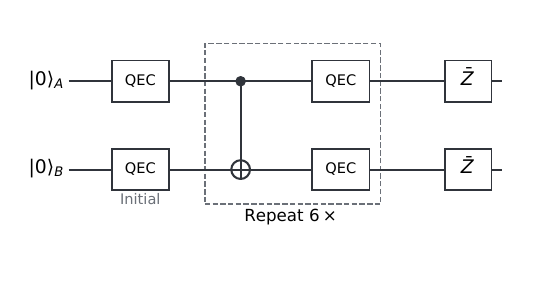}
\caption{Fast transversal CNOT circuit. Two unrotated $d=5$ surface code patches are initialized in logical $\lvert0\rangle$. An initial quantum error correction (QEC) round is followed by six repetitions of a transversal CNOT from A (control) to B (target) and one QEC round. There is one QEC round between successive gates and seven rounds in total. Both patches are finally measured in logical $Z$. The dashed box denotes the repeated circuit block. The noise model is SI1000 with $p=0.001$. Decoder results are listed in Table~\ref{tab:gates-main}.}
\label{fig:gates}
\end{figure}

We simulate the repeated transversal-CNOT circuit of Ref.~\cite{serraperalta2026lom} on two unrotated $d=5$ surface code patches [Fig.~\ref{fig:gates}]. The circuit applies six CNOTs in the same direction, with logical-$Z$ initialization and readout. One quantum error correction (QEC) round follows each gate, giving seven rounds including the initial round. We use SI1000 noise at $p=0.001$.

All decoders receive the same 20,000 circuit samples. We measure the circuit failure probability $P_{\mathrm{fail}}$, the probability that either independent logical observable is incorrect at the end of the six-gate circuit. LOM and SMP+LOM use the same matching backend; the latter adds the motif-based weight updates. This evaluation reuses previously sampled data and has no independent validation set.

\begin{table}[!htbp]
\caption{Decoder comparison for the six-CNOT circuit in Fig.~\ref{fig:gates}. Each decoder uses the same 20,000 circuit samples. $P_{\mathrm{fail}}$ is the circuit failure probability: either independent logical observable is incorrect. Probabilities and marginal Wilson 95\% intervals are in percent. $R$ is the relative reduction from LOM+MWPM, expressed as a percentage. LOM and SMP+LOM use MWPM. BP10 uses at most 10 belief propagation iterations. It outputs its prediction on convergence; otherwise it uses LOM+MWPM with updated probabilities. Table~\ref{tab:gates} also includes BP100, which allows at most 100 iterations.}
\label{tab:gates-main}
\centering
\footnotesize
\setlength{\tabcolsep}{3pt}
\begin{ruledtabular}
\begin{tabular}{lrcr}
Decoder & $P_{\mathrm{fail}}$ (\%) & 95\% interval & $R$ (\%) \\
\hline
\GateMainRows
\end{tabular}
\end{ruledtabular}
\end{table}

SMP reduces the circuit failure probability from 0.535\% to 0.345\%, a relative reduction of 35.5\% [Table~\ref{tab:gates-main}]. On the shared samples, it corrects 49 shots that LOM fails and introduces 11 new failures. This result supports using the complete motif to set weights even when the final correction is chosen in separate observable-specific graphs.

To compare this gain with another use of fault correlations, we also evaluate belief-propagation-assisted LOM. BP10 and BP100 allow at most 10 and 100 BP iterations, respectively. Their observed failure probabilities are 0.320\% and 0.260\%, both below the SMP value [Table~\ref{tab:gates}; paired comparisons in Appendix~\ref{app:gates}]. Establishing the corresponding accuracy--runtime trade-off requires execution-time measurements, which were not collected in this evaluation.

\section{Discussion}
\label{sec:discussion}
In this work, we developed Syndrome Motif Projection (SMP) as a general framework for handling hyperedge faults. We have applied SMP to surface codes, color codes, and logical gate circuits, where it consistently outperforms the corresponding baseline matching decoders.

This opens several directions beyond the settings studied here. Lattice surgery provides one example. Merge and split operations change the stabilizer measurements and produce new fault motifs \cite{rotated_horsman2012surface,circuit_chamberland2022universal}. For example, crosstalk and leakage can generate correlations across space or time \cite{correlated_crosstalk,liu2025parallelcrosstalk,zhou2025crosstalk,mcewen2021leakage}. Leakage-reduction circuits and flag-based extraction can also produce characteristic local fault patterns \cite{mcewen2021leakage,he2025leakagesuppression,chamberland2020flags}.

The $O(d^2r)$ preprocessing cost derived in Sec.~\ref{sec:workflow} supports efficient implementation. Only activated motifs require weight updates, so the additional overhead decreases in the low-noise regime. SMP modifies graph weights while leaving the matching solver unchanged. It can therefore be integrated with existing parallel and sliding-window decoding architectures \cite{fowler2012practical,sparse_blossom,tan2023parallel,fusion_blossom,ziad2025lcd,skoric2023window}. The same motif evidence could also guide cluster growth in weighted union-find decoders, extending the idea beyond matching \cite{delfosse2021unionfind,huang2020weighteduf,circuit_2023improved}. An important next step is to implement SMP in parallel hardware and measure its end-to-end decoding latency. Recent real-time decoding demonstrations provide suitable architectures for testing SMP at microsecond-scale latency \cite{caune2026realtime,yang2026fpgadecoder}.

\begin{acknowledgments}
The authors acknowledge support from the National Natural Science Foundation of China (Grants No.~12174215 and No.~12374473) and the Quantum Science and Technology-National Science and Technology Major Project (No.~2021ZD0300705).
\end{acknowledgments}

\appendix
\section{Local-posterior reference implementation}
\label{app:local-posterior}

This appendix describes the local posterior implementation used for comparison with SMP. It estimates the evidence in Eq.~\eqref{eq:smp_local_configuration_evidence} from allowed local edge configurations of matching graph edges. It then uses the normalized posterior to interpolate cluster costs.

\subsection{Candidate hypotheses and local regions}

The implementation starts from a geometry-based predecoder catalog. Candidates with intersecting detector supports are grouped into connected components. The union of their supports defines a seed region. A greedy pass examines matching edges connected to the seed detectors in increasing order of their prior weight. It can join an active seed detector to another active detector or to a boundary. The touched detector vertices enlarge the region to $\Omega$. This expansion only defines the local region; its edges are not assumed to have occurred. Seed regions or expanded regions containing more than 20 detectors are skipped.

Each region contains an independent-edge branch and one branch for each retained candidate. A candidate can contain several hyperedge members and explicitly included matching edges. Its prior odds are the product of the member odds and the included edge odds. For a single hyperedge this factor is $P_H$. Member priors are obtained from matching DEM support records when available; the implementation otherwise uses its motif type prior. Duplicate descriptions with the same support, target edges, and recorded logical-observable flips are merged by adding their relative prior weights. This is the convention of this comparison implementation. It differs from the independent-mechanism parity aggregation in Eq.~\eqref{eq:smp_combined_prior}.

\subsection{Probabilities of local edge configurations}

For the independent-edge branch, the required detector set is $Y\cap\Omega$. For a candidate $h$, it is $(Y\triangle S_h)\cap\Omega$. In the latter branch, edges touching $(Y\cap S_h)\cap\Omega$ are forbidden. This prevents the residual edge configuration from reusing an observed detector already attributed to the candidate. For a composite candidate, its full XOR support replaces $S_h$ in these rules.

The routine collects matching edges connected to at least one required detector and removes edges touching a forbidden detector. An allowed local edge configuration uses each required detector exactly once. Other endpoints are not constrained, except by the forbidden-detector rule. These rules determine the configurations included in the local score.

At most eight distinct local edge configurations are generated for each branch. A greedy completion first chooses the unassigned required detector with the fewest available options. It selects the available edge with the largest prior odds and continues until each required detector has been assigned an edge. Further completions force one starting edge before applying the same procedure. Edges joining two required detectors are tried first. Duplicate edge sets are counted once. A failed completion contributes no mass. The empty required set has unit relative mass.

Each allowed local edge configuration contributes the product of its edge odds, as in Eq.~\eqref{eq:smp_local_edge_mass}. The implementation sums these products using log-sum-exp. The candidate branch multiplies this residual mass by its prior odds. Branches with no allowed local edge configuration are excluded. Extending the normalization in Eq.~\eqref{eq:smp_local_posterior_share} to several competing candidates in the same region gives
\begin{equation}
 \alpha_h
 =\frac{e^{-c_h}}{e^{-c_I}+\sum_{h'}e^{-c_{h'}}}.
 \label{eq:local_posterior_competition}
\end{equation}
The sum includes all retained candidate branches in that region. These branches are treated as mutually exclusive alternatives, including any composite branches. Their probabilities are therefore normalized within a restricted hypothesis model.

\subsection{Weight updates and comparison settings}

As an implementation safeguard, a negative target cluster cost can be set to zero. This nonnegative floor is inactive for $p_h<1/2$ and does not alter the cost definition in Eq.~\eqref{eq:smp_posterior_cost_reference} in the noise regime considered here.

The tested implementation uses posterior interpolation with unit interpolation power and permits only decreases from the original cluster cost. It distributes each decrease in proportion to the original target edge weights, as in Eq.~\eqref{eq:smp_posterior_normalized_weight}. Contributions from different candidate branches are accumulated on their target edges. The final weights have a zero floor. Matching edges used only to explain the residual syndrome receive no posterior update. The original syndrome is decoded with standard MWPM after all updates. This regional normalization and accumulation differ from SMP's per-edge competition in Eq.~\eqref{eq:smp_overlap_weight}.

The comparison entry point is \texttt{run\_posterior\_comparison.py} in the accompanying \texttt{posterior\_matching} experiments. It calls the shared local posterior implementation in \texttt{decoder\_v2\_soft\_and\_hard\_information}. The historical mode name \texttt{cluster\_factor\_graph\_exact} is retained in the code. The active routine nevertheless uses the bounded sum over greedily generated local edge configurations described above; the legacy exact-state budget is not used by that routine.

Table~\ref{tab:local-posterior-comparison} summarizes the saved comparisons. All three methods use the same seeded samples, circuit, decomposed DEM, and standard MWPM backend at each point. The circuits are unrotated surface code memories in logical $Z$, with $r=d$. Depolarizing noise follows Clifford gates and precedes each extraction round on data qubits. Measurement errors and reset errors also occur with probability $p$. The candidate catalogs differ: the local posterior version uses the geometry-based predecoder, whereas the SMP approximation extracts three- and four-detector supports from the DEM. The comparisons therefore evaluate the complete implementations.

\begin{table}[tb]
\caption{Saved local posterior comparisons on shared surface code samples. Entries are logical failure counts. All methods use standard MWPM. The two evidence implementations also differ in their candidate catalogs.}
\label{tab:local-posterior-comparison}
\centering\small
\begin{tabular}{ccrrrr}
\toprule
$d=r$ & $p$ & Shots & MWPM & Local posterior & SMP \\
\midrule
3 & 0.008 & 10,000 & 590 & 552 & 543 \\
5 & 0.010 & 2,000 & 211 & 195 & 202 \\
7 & 0.008 & 20,000 & 1242 & 1082 & 1043 \\
\bottomrule
\end{tabular}
\end{table}

The local posterior and SMP predictions have unadjusted two-sided exact McNemar $p$ values of 0.3742, 0.4101, and 0.0412 for the three rows. At $d=7$, SMP corrects 193 local posterior failures and introduces 154 failures, reducing the observed failure probability from 5.410\% to 5.215\%. These comparisons show competitive decoding accuracy for the simpler SMP evidence model in the tested settings. They do not establish equivalence, numerical accuracy of its posterior estimates, or a gain attributable solely to the likelihood approximation.

\section{Willow data, decoder settings, and local evidence}
\label{app:willow-decoders}
This appendix specifies the Willow data and decoder settings, the per-round fitting procedure, and the local score used in the surface code benchmarks. These details connect the comparisons in Sec.~\ref{sec:surface} to the implementation.

We use patches \texttt{d3\_at\_q6\_7}, \texttt{d5\_at\_q6\_5}, and \texttt{d7\_at\_q6\_7}. Each combination of distance, basis, round count, and prior uses its corresponding released DEM. The released RL prior was optimized jointly for $d=3$ and $d=5$ patches using 13-cycle calibration data \cite{willow2025,sivak2024priors}. We reuse that prior and do not retrain it.

``Correlated matching (Google released)'' and ``Harmony-51'' refer to Google's saved predictions for the indicated prior. ``Correlated MWPM'' refers to our evaluation with PyMatching 2.3.1 and correlations enabled \cite{higgott2022pymatching}. SMP supplies syndrome-dependent weights using the same DEM. SMP with standard and correlated MWPM use the same prepared proposals. The original conditional correlation rules of the correlated backend are retained; they are not re-estimated from the SMP scores.

Belief matching uses \texttt{beliefmatching} 0.2.0 on the unmodified full Google DEM. Its sum-product BP has at most ten iterations and uses one OpenMP thread. The archived $d=7$ comparison covers $r=13$. The SI1000 logical failure probabilities are 2.788\% in X and 2.294\% in Z. The corresponding RL values are 2.480\% and 1.950\%. Each basis contains 50,000 shots, giving the pooled entries in Table~\ref{tab:willow-selected}. Missing round counts are not treated as zero-error measurements.

The per-round reductions also depend on logical basis. At $d=7$, the largest reductions occur in logical $Z$ under the SI1000 prior. SMP lowers the MWPM rate from 0.3906\% to 0.2968\% per round, a 24.0\% reduction. It lowers the correlated MWPM rate from 0.2155\% to 0.1961\%, a 9.0\% reduction. These basis-resolved values differ from the X/Z averages plotted in Figs.~\ref{fig:willow-mwpm} and \ref{fig:willow-correlated}.

For the per-round fits, we regress $\log[1-2P_{\mathrm L}(r)]$ against $r$. The slope determines $\epsilon$ through Eq.~\eqref{eq:per-round-fit}, and the intercept determines $F_0$. Following the basis-resolved fitting procedure in Sec.~\ref{sec:surface}, a parametric bootstrap resamples each of the four binomial counts for each basis and repeats the fit 5,000 times. The plotted intervals are the 2.5th and 97.5th percentiles of the averaged rates. These marginal intervals are not paired confidence intervals for a difference between decoders.

For the surface code SMP implementation, independent mechanisms with the same detector support $S_h$ and logical effect $\bm\ell_h$ are grouped into one candidate $h$. Let $\mathcal R_h$ denote this group. Its prior is the probability of an odd number of these faults \cite{stimdemformat}:
\begin{equation}
 p_h=\frac{1}{2}\left[1-\prod_{r\in\mathcal R_h}(1-2p_r)\right].
 \label{eq:smp_combined_prior}
\end{equation}
Mechanisms with different logical effects remain separate candidates.

The surface code score reduces the evidence for a candidate when some expected events are missing. We count these events as $m_h=k_h-n_h$ and estimate their cancellation by the two background channels of Sec.~\ref{sec:smp_approximation}. The probability that exactly one channel occurs is
\begin{equation}
 q=p_{\mathrm{iid}}(1-p_{\mathrm{meas}})+(1-p_{\mathrm{iid}})p_{\mathrm{meas}}.
 \label{eq:surface-hidden-q}
\end{equation}
The score assigns a factor $q/(1-q)$ to each missing event and a fixed unit penalty in the log score to three-detector candidates. For an activated motif, this gives
\begin{equation}
 \begin{split}
 \log\eta_h={}&\log P_H-\log(P_{\mathrm{iid}}P_m)\\
 &-\mathbf1_{k_h=3}-m_h\log\frac{1-q}{q}.
 \end{split}
 \label{eq:surface-implementation-score}
\end{equation}
The indicator $\mathbf1_{k_h=3}$ equals one for a three-detector support and zero otherwise. This penalty is an implementation convention, not a consequence of the two-channel derivation. For a missing event, the implementation uses $q/(1-q)$, whereas Eq.~\eqref{eq:smp_hidden_weight} compares exactly one background channel with neither. Since
\begin{equation}
 \frac{q}{1-q}=\frac{P_{\mathrm{iid}}+P_m}{1+P_{\mathrm{iid}}P_m},
\end{equation}
the two expressions agree to leading order at low noise. In the Willow benchmark, $p_{\mathrm{iid}}=0.002$ and $p_{\mathrm{meas}}=0.005$ are fixed scoring parameters. They are not fitted edge probabilities. The candidate probability $p_h$ comes from the selected DEM through Eq.~\eqref{eq:smp_combined_prior}. In the uniform-noise surface code simulations, both scoring parameters equal the nominal $p$.

Motif activation follows Sec.~\ref{sec:workflow}. There is no recursive motif growth, virtual syndrome flip, or merging of observed motifs into larger clusters in these benchmarks. Duplicate proposals for the same physical representation are combined before the shared edge competition. The local derivation in Sec.~\ref{sec:smp} motivates the weighting structure, while Eq.~\eqref{eq:surface-implementation-score} specifies the benchmark evidence.

\section{Surface code depolarizing noise scan}
\label{app:surface-scan}
This appendix gives the circuit noise specification and the noise scan behind the fixed-noise comparison in Sec.~\ref{sec:surface}. The scan tests the persistence of the gain over the sampled range.

\begin{figure*}[tp]
\centering
\includegraphics[width=0.95\textwidth]{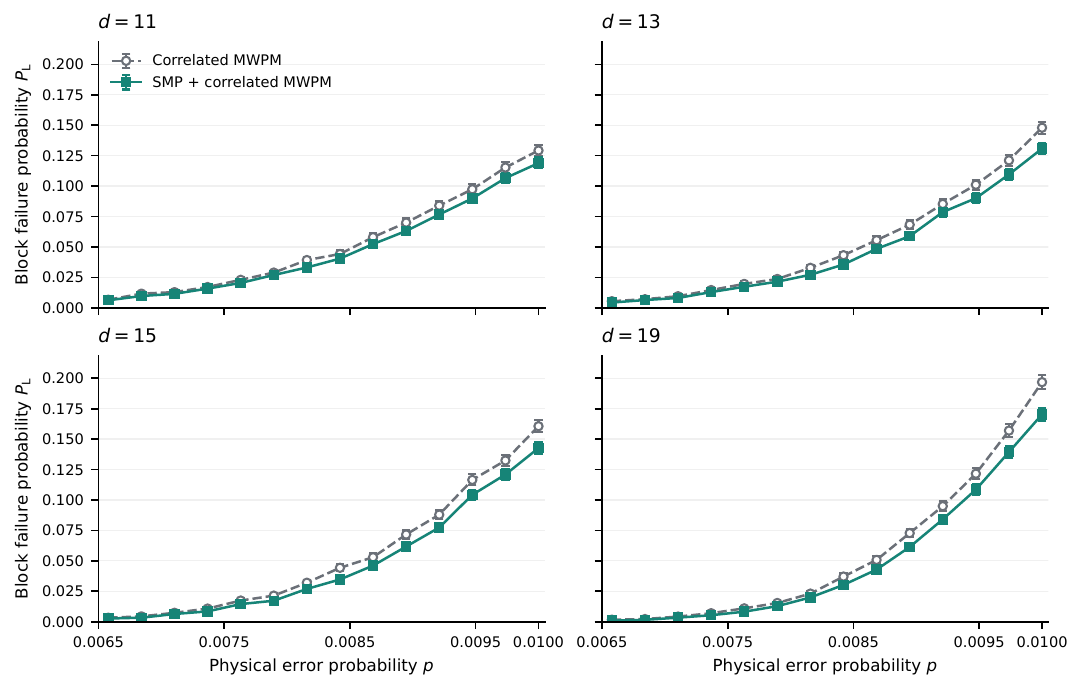}
\caption{Surface code logical failure probabilities per memory experiment in the interval $0.0065\leq p\leq0.01$ under uniform circuit-level depolarizing noise. Panels show unrotated logical-$Z$ memories with $d=11,13,15,19$ and $r=d$. Both axes are linear; all panels share the same vertical scale with a zero origin. The horizontal coordinate is a probability, not a percentage. Gray dashed curves with open circles show correlated MWPM; teal solid curves with filled squares show SMP-correlated MWPM. Every original point in this interval is included. Each point uses 20,000 shared shots. Bars are marginal Wilson 95\% intervals; connecting lines guide the eye. This is a restricted view of the original 20-point scan, with unchanged samples and statistics. Figure~\ref{fig:depolarizing} shows a fixed-$p$ slice.}
\label{fig:surface-scan}
\end{figure*}

The complete dataset contains 20 noise strengths from 0.005 to 0.01 at each of $d=11,13,15,19$. Figure~\ref{fig:surface-scan} displays the 14 sampled strengths in $0.0065\leq p\leq0.01$. The first displayed point is $p=0.006578947\ldots$; no point is interpolated at 0.0065. The full 20-point scan is retained in the accompanying CSV. The circuits, noise model, and shared samples are those of Sec.~\ref{sec:surface}.

Low-noise points at large distance contain few failures, so their relative gains have substantial sampling uncertainty. The plot retains every sampled point in the displayed interval and its marginal Wilson interval. We use the scan to test persistence over a finite range, not to extract a threshold or establish monotonic improvement with distance. This dataset is distinct from rotated-code simulations with $r=3d$.

\section{Color code projection and paired results}
\label{app:color}
This appendix defines the uniform-hidden score and activation criteria used with the cost rule in Sec.~\ref{sec:color-problem}. It also records the projection checks, replay provenance, and complete paired results for Sec.~\ref{sec:color}.

The Middle-out circuits of Sec.~\ref{sec:color-results} are passed to Chromobius without conversion from CX to CZ \cite{gidney2023chromobius}. At every point, the first 1,000 shots were also checked against the local unmodified M\"obius path. Their logical predictions agree.

The color code score compares explanations of the observed part $O_h$ of the motif with and without candidate $h$. Both explanations require an estimate of detector flips caused by other faults. Candidate mechanisms are grouped only when both detector support $S_h$ and logical-observable flip vector $\bm\ell_h$ agree. Let $\mathcal R_h$ contain the mechanisms grouped into $h$. Excluding this group avoids counting the candidate as part of its own background. For detector $v$, the odd-parity formula in Eq.~\eqref{eq:smp_combined_prior} gives the background flip probability
\begin{equation}
 q_{v|\neg h}=\frac12\left[1-\prod_{\substack{r\in\mathcal R\setminus\mathcal R_h\\ v\in S_r}}(1-2p_r)\right].
 \label{eq:color-marginal}
\end{equation}
The individual marginals are exact under the DEM prior. To obtain a score that depends only on the observed event count, we approximate the background bits as independent flips with a common probability equal to their mean:
\begin{equation}
 \bar q_h=\frac1{k_h}\sum_{v\in S_h}q_{v|\neg h}.
\end{equation}
This approximation averages over spatial variation and neglects correlations. Without $h$, the background must produce the $n_h$ observed events. With $h$, it must instead cancel the $k_h-n_h$ missing events. Comparing these two possibilities and including the prior odds gives the log evidence ratio
\begin{equation}
 \begin{aligned}
 a_h(n_h)&=\log P_H
 +(2n_h-k_h)\log\frac{1-\bar q_h}{\bar q_h},\\
 \eta_h&=e^{a_h(n_h)}.
 \end{aligned}
 \label{eq:color-count}
\end{equation}
This is the evidence ratio in Eq.~\eqref{eq:smp_local_posterior_share} with $c_I(Y)-c_h(Y)=a_h(n_h)$ under the uniform-hidden approximation. The count $n_h$ refers to physical detectors before the M\"obius duplication. We precompute Eq.~\eqref{eq:color-count} for $n_h=0,\ldots,k_h$, giving $k_h+1$ scores per motif. The background probability $\bar q_h$ differs from the relative cancellation weight $P_{\mathrm{hidden}}$ in Eq.~\eqref{eq:smp_hidden_weight}.

We consider support sizes three through seven. Three-detector candidates activate when at least two events are observed. Larger candidates require at least three. Projection uses the upstream M\"obius construction and requires the projected detector support $\iota(S_h)$ and logical-observable flip vector $\bm\ell_h$ under forced lifting. Empty clusters and clusters that fail the parity checks are excluded. Retained clusters have target cost $2W_{C,h}$ and satisfy $0\leq2W_{C,h}<W_{E,h}$. The same cluster-cost convention applies to composite faults.

Table~\ref{tab:color} reports the current uniform-hidden replay on saved extension samples generated independently of the earlier confirmation set. The archived replay verification matched the sample, DEM, and baseline prediction hashes to each replay record. It reproduced the reported error probabilities and paired statistics from the saved predictions. The uniform-hidden and preceding count implementations give identical predictions at all six $p=0.001$ points. At $p=0.003$, their predictions differ slightly. We use the uniform-hidden version consistently.

The archived single-mechanism audit for the preceding count implementation covered 439,124 DEM mechanisms across the twelve cases, with no logical failures. It checks the retained projection and lifting construction. It is separate from the archived replay verification and does not certify every later evidence approximation or multiple-fault configuration.

\begin{table*}[tp]
\caption{Color code logical failure probabilities per memory experiment for the current uniform-hidden implementation. Each row uses 100,000 shared shots with $r=2d$. The $P_{\mathrm L}$ and relative reduction $R$ columns are percentages. The baseline is official Chromobius. The last column gives unadjusted two-sided exact McNemar $p$ values. These statistical $p$ values differ from the noise strength $p$ in the first column. Figure~\ref{fig:color} shows low-noise uncertainty intervals.}
\label{tab:color}
\centering\small
\begin{tabular}{cccrrrr}
\toprule
$p$ & $d$ & Basis & $P_{\mathrm{L,Chromobius}}$ (\%) & $P_{\mathrm{L,SMP}}$ (\%) & $R$ (\%) & Paired $p$ value \\
\midrule
\ColorRows
\bottomrule
\end{tabular}
\end{table*}

\section{SMP projection to LOM and comparison settings}
\label{app:gates}
This appendix gives the scoring and implementation settings for the LOM update in Eq.~\eqref{eq:lom-discount}. It also specifies the decoder comparisons and the saved-DEM checks supporting Fig.~\ref{fig:cnot-motif}.

The gate circuit and LOM construction follow Ref.~\cite{serraperalta2026lom}. We use the saved circuit rather than regenerating it from a seed, because equivalent detector orderings can differ between generator environments. The failure criterion in Sec.~\ref{sec:gates-results} uses the two independent logical observables and excludes the redundant third product observable.

The present benchmark concerns transversal CNOTs. More general Clifford operations can use folded surface codes and fold-transversal constructions \cite{moussa2016folded,breuckmann2024fold}. A universal transversal gate set is excluded for a fixed code that detects arbitrary single-subsystem errors \cite{eastin2009restrictions}, and constant-depth geometrically local gates obey further restrictions \cite{bravyi2013classification}. Universal constructions therefore use additional ingredients, such as magic state distillation, recovery steps that restore the code space, or gauge fixing \cite{bravyi2005universal,paetznick2013universal,bombin2015gauge}. Generalized lattice surgery also allows a non-Clifford gate implemented transversally in one code to be teleported to another code \cite{wang2026gateteleportation}. These extensions are outside the six-CNOT test reported here.

The implementation sets negative proposed edge costs to zero when the reduction exceeds the original projected cluster cost. This safeguard applies after the proportional allocation in Sec.~\ref{sec:smp_general}.

When constructing proposals for Eq.~\eqref{eq:lom-discount}, an unmatched full support, an empty projection, or an absent parity-preserving native cluster produces no update. The reference graph used to compute $W_{\mathrm{full},h}$ retains the archived splitting convention. It omits components that cannot be represented as matching edges. Sampling still uses the complete physical circuit.

The score in Eq.~\eqref{eq:surface-implementation-score} uses $p_{\mathrm{iid}}=p_{\mathrm{meas}}=0.001$. Support sizes 3, 4, 5, 6, and 8 can contribute, with at least three observed detectors required. The full-syndrome cost affects only the proposed edge costs. We apply Eq.~\eqref{eq:smp_overlap_weight} without rescaling the log evidence.

The software is \texttt{lomatching} 0.2.2, Stim 1.15.0, and PyMatching 2.3.1. BP-assisted LOM uses min-sum BP, scaling factor one, parallel schedule, and at most 10 or 100 iterations. If BP converges, its estimated faults supply the logical prediction directly. Otherwise, the BP probabilities initialize the LOM matching problems. The comparison reuses the saved BP predictions on the same shots and contains no timing measurements.

\begin{table}[!htbp]
\caption{Decoder comparison, including BP100, for the six-CNOT circuit in Fig.~\ref{fig:gates}, using the same 20,000 circuit samples. Circuit failure probabilities and marginal Wilson 95\% intervals are in percent. $R$ is the relative reduction from LOM+MWPM, expressed as a percentage.}
\label{tab:gates}
\centering\footnotesize\setlength{\tabcolsep}{3pt}
\begin{tabular}{lrcr}
\toprule
Decoder & $P_{\mathrm{fail}}$ (\%) & 95\% interval & $R$ (\%) \\
\midrule
\GateRows
\bottomrule
\end{tabular}
\end{table}

The paired comparison of SMP and LOM reported in Sec.~\ref{sec:gates-results} gives a two-sided exact McNemar $p=7.56\times10^{-7}$.

The BP comparisons do not establish equal performance. Relative to BP10, SMP corrects 37 failures and introduces 42, giving $p=0.653$. Relative to BP100, the corresponding counts are 28 and 45, giving $p=0.0604$. These latter tests are unadjusted. Failure to reject equal marginal error rates does not demonstrate equivalence. These comparisons reuse the existing samples, without a new independent validation set.

The three-point example in Fig.~\ref{fig:cnot-motif} can be verified without sampling. The representative $X_{29}$ fault occurs immediately after reset at instruction 331, after 18 ticks. Its Pauli support remains on ancilla 29, and it flips only $m_{94}$. The grouped raw DEM mechanism with support $\{D_{94},D_{134},D_{174}\}$ has probability $0.00757149$; this is not the probability of the representative location alone. Its projections are $\{D_{94},D_{174}\}$ in $G_0$ and $\{D_{134},D_{174}\}$ in $G_1$. Each pair maps to local indices $\{26,46\}$ in its respective graph. Both saved LOM DEMs contain the ordinary instruction $\texttt{error}(0.00829355)\,D_{26}D_{46}$, with no correlated component separator or target logical-observable flip label. The different probabilities reflect marginalization with other physical mechanisms. The corresponding prior edge weight is approximately $4.784$. This is a check of the stored graph representation; the example is not itself a logical failure. The local reproduction script records the full supports, coordinate maps, native edge attributes, and representative circuit location.

Archived checks compared mutable updates with fresh matchers and verified projected parity. The full-syndrome SMP implementation had zero failures when enumerating all 11,756 single DEM mechanisms. For this manuscript, we recounted all 20,000 saved shots and recovered every entry in Table~\ref{tab:gates}. Neither check proves fault tolerance for arbitrary gate circuits.

\bibliographystyle{apsrev4-2}
\bibliography{refnew}
\end{document}